\documentclass[11pt]{article}
\usepackage{amsfonts}
\usepackage{fullpage}
\usepackage{graphicx}
\usepackage{amsmath}
\usepackage{amsbsy}
\usepackage{amssymb}
\usepackage{mathtools}
\usepackage{mathrsfs}
\usepackage{tabularx}
\usepackage{indentfirst} 
\usepackage{ifpdf}
\usepackage{subcaption,graphicx}
\usepackage{algorithm} 
\usepackage[noend]{algpseudocode}
\usepackage{algorithmicx}
\usepackage{algpseudocode}
\usepackage{setspace}
\usepackage{lscape}
\usepackage{pdflscape}
\usepackage{rotating}
\usepackage{enumitem}
\usepackage{empheq}
\usepackage[title]{appendix}
\algdef{SE}[DOWHILE]{Do}{doWhile}{\algorithmicdo}[1]{\algorithmicwhile\ #1}%

\usepackage{color}
\RequirePackage{xcolor}
\usepackage{mathtools}
\usepackage[utf8]{inputenc}
\usepackage{geometry}
\usepackage[export]{adjustbox}
\usepackage{float}
\usepackage[english]{babel}
\usepackage{xr}
\usepackage{booktabs}
\usepackage{multirow}
\usepackage{rotating}
\usepackage{alphalph,etoolbox}
\usepackage[english]{babel}
\usepackage{hyperref}  
\usepackage{breakurl}
\addto\captionsenglish{
	\renewcommand{\contentsname}%
	{Appendices}%
}
\patchcmd{\subequations}{\alph{equation}}{\alphalph{\value{equation}}}{}{}

\algcblockdefx[Then]{If}{Then}{EndThen}{$)$ $\{$}{$\}$}
\algcblockdefx[Else]{Then}{Else}{EndElse}{$\}$ \textbf{else} $\{$}{$\}$}

\date{}
\begin{document}
	\renewcommand{\thepage}{\roman{page}}
	\setcounter{page}{0}
	
	\title{\textbf{Call Window Scheduling for Freight Rail Engineers}\footnote{This work was supported in part by Care Systems, Inc. under grant \# 30-2110-1412.}}
	\author{Jia Guo\footnote{Corresponding author: jia.guo@utexas.edu; postal address: Room 111, H20, Research Centre for Integrated Transport Innovation (rCITI), School of Civil and Environmental Engineering, the University of New South Wales, Kensington, NSW 2033, Australia.} \\
		Jonathan F. Bard}
	\maketitle
	\thispagestyle{empty}
	
	\begin{center}
		\text{Operations Research and Industrial Engineering} \\
		\text{Cockrell School of Engineering}\\
		\text{The University of Texas} \\
		\text{Austin, Texas 78712, USA}\\
	\end{center}
	\vspace{0.5in}
	
	\newgeometry{top=20mm, bottom=20mm, left=20mm, right=20mm}
	
	\newpage
	\begin{center}\textbf{Call Window Scheduling for Freight Rail Engineers}\\
	\end{center}

\begin{abstract}
This study investigates a new approach for scheduling freight rail engineers based on call windows under the Hours of Service (HOS) regulations. Unlike planned trip assignments, call windows specify a time interval during which an engineer may be required to start work, thus providing greater flexibility to handle uncertain trip demand while offering drivers more predictable off-duty periods. Currently in the U.S., all major freight operators require drivers to be available 24/7 outside of mandatory rest periods, raising concerns over workforce fatigue and safety. We formalize the call-window scheduling problem and propose two solution approaches: a Set-covering-type optimization model and a Direct Algorithm. Both aim to maximize demand coverage while minimizing the number of engineers subject to HOS feasibility. Computational experiments for 2-city and 3-city instances with varying call window lengths, maximum delay allowances, and whether to allow deadhead trips show that the Set-covering-type model yields higher demand undercoverage (94.89\% and 91.09\% for 2-city and 3-city instances, respectively) than the Direct Algorithm (93.96\% and 71.83\%). It also offers greater rest opportunities and reduced delays. Sensitivity analysis reveals that the upper limit on engineer availability significantly affects all key performance metrics.
\end{abstract}


\renewcommand{\thepage}{\arabic{page}}
\setcounter{page}{1}
{\tiny }

\section{Introduction} \label{sec_introduction}

\noindent Freight railways operate in a highly uncertain, 24/7 environment in the U.S. Demand for train movements can vary by time of day, day of week and season, and where strict safety and fatigue rules constrain how long drivers may remain on duty. Traditional personnel scheduling assigns drivers to specific trips. Such rigid assignments, however, can force operators to maintain a large number of drivers effectively “on call” or available around the clock to cover unpredictable schedule variations.  This leads to high labor costs, increased idle time, and fatigue risk. The study on railway crew scheduling by Heil et al. (2020) documents both the complexity of these issues and the trade-offs among demand coverage, crew utilization and Hours of Service (HOS) regulations imposed by federal law (Public Law 110-432).  These issues do not arise in passenger train crew scheduling because time tables are determined months in advance. This makes it is possible to derive high quality driver assignments with known days on and days off for a given planning horizon.

\indent One way to introduce controlled flexibility is to use call windows rather than directly assigning drivers to upcoming trips. Call windows specify a time interval during which a driver may be required to start a duty. The exact trip assignment is decided later when more demand information is available, such as when departure times are finalized or when delays due to track congestion are cleared.  Call window scheduling approaches have been studied in other personnel scheduling domains as a way to address demand uncertainty while reducing over-coverage and improving responsiveness (Gawas et al. 2023). 

\indent The need for formal call window scheduling in freight train operations is reinforced by regulatory and safety constraints. HOS laws and rules for train employees (which mandate limits on consecutive work hours, minimum rest periods and statutory off-duty periods) restrict how often and how long drivers may be on call, so any call window policy must guarantee feasible schedules under these rules. Incorporating HOS constraints is therefore critical to maintain safety and legal compliance while seeking efficiency gains. 

\indent At present, all freight operators in the U.S. rely on informal 24/7 availability practices that implicitly keep a pool of drivers available to respond at short notice; sized schedules are not used. This practice creates three problems. First, it inflates the effective workforce required to maintain a target level of demand coverage, increasing labor cost and administrative burden. Second, it increases the risk of fatigue or regulatory violations if rest and duty limits are not carefully monitored. Third, it disrupts normal living, making it difficult to have a satisfactory work-life balance.  Introducing call windows promises a middle ground: by specifying start windows instead of fixed assignments, operators can (i) postpone the final matching of drivers to trips until after uncertain demand is realized, (ii) reduce the number of drivers who must be continuously available, and (iii) increase the ability to respect HOS constraints while maintaining demand coverage.

\indent This paper studies call window scheduling for freight train drivers traveling between two or three cities (the standard practice in the U.S.), and it makes the following four main contributions. 

\begin{enumerate}
	\item We formalize the call window scheduling problem for freight engineers, explicitly modeling HOS constraints and the operational requirement that drivers may only start a duty within their assigned call window. We are the first to propose such an approach for the freight rail industry.
	\item We develop two solution procedures based on: (i) a Set-covering-type integer programming model that selects a minimum (or cost-effective) set of call window schedules to cover demand, and (ii) a Direct Algorithm that constructs and assigns call windows to drivers in an efficient manner.
	\item We perform computational experiments on instances of realistic size to compare solution quality and runtimes associated with the two algorithms. Solution robustness in the face of uncertain rail schedules is demonstrated through simulation.
	\item We conduct an extensive sensitivity analysis to investigate the impact of maximum labor resources on demand coverage and other cost metrics.
\end{enumerate}

\indent The remainder of the paper is organized as follows. Section \ref{sec_literature} reviews the relevant literature on rail/road personnel scheduling and call window scheduling in other service domains. Section \ref{sec_probdes} formally defines the call window scheduling problem and outlines the feasibility conditions imposed by the HOS regulations. Sections \ref{sec_method} and \ref{sec_method2} present our two solution procedures: the Set-covering-type formulation and the Direct Algorithm, including modeling details and solution methodology. Sections \ref{sec_compexp_method1} and \ref{sec_compexp_method2} highlight the computational results for the two procedures, respectively, while Section \ref{sec_compexp_compare} compares their performance. We close with a discussion of managerial insights, study limitations, and potential extensions for future research in Section \ref{sec_conclusion}.

\section{Literature Review} \label{sec_literature}
\noindent The literature on personnel scheduling is vast. To maintain focus, we limit the discussion to the most relevant research. In Section \ref{sec_literature_train}, we review personnel scheduling problems for rail and road engineers/drivers, while Section \ref{sec_literature_window} highlights the prior work on call window scheduling in freight rail and other fields. In Section \ref{sec_literature_gap}, we identify the research gap and novel contributions of this study.

\subsection{Personnel Scheduling for Rail and Road Engineers/Drivers} \label{sec_literature_train}
\noindent This subsection provides a review on the personnel scheduling problems for rail and road engineers/drivers in chronological order. Vaidyanathan et al. (2007) investigated a railway crew scheduling problem for freight rail in North America using a planning horizon of 1–4 weeks. A successive constraint generation (SCG) algorithm and a quadratic cost perturbation (QCP) algorithm were proposed to assign engineers to scheduled trips between two terminals. The tasks included driving, deadheading by train or taxi to the home base, and rest. Their objective was to minimize a weighted combination of the number of engineers, deadhead costs, crew rest and train delays. Trip coverage, engineer qualification, minimum rest time and maximum consecutive on duty time constituted the hard constraints. Given a 1-hour runtime limit, SCG and QCP provided solutions to  a 326-trip instance with an optimality gap of 0.07\% and 0.03\%, respectively. Our problem is similar to theirs but includes many more constraints imposed by the HOS regulations, as modified in the years subsequent to 2007.

\indent In the study by Nishi et al. (2011), a daily crew scheduling problem for Japaneses freight carriers was investigated. Column generation and dual inequalities from a Dantzig–Wolfe decomposition were applied to minimize the number of engineers. Each engineer started and ended at the same home base and all trips were required to be covered. Constraints on the minimum rest time and maximum working time for each engineer had to be satisfied. For random instances with up to 1389 trips, the proposed algorithm found solutions with costs that were 41.15\% lower than those from a branch-and-bound commercial code.

\indent Jütte et al. (2011) investigated a freight crew scheduling problem for the German company DB Schenker with a planning horizon of one week. Their objective was to minimize the weighted sum of penalty costs consisting of consecutive duty durations of more than 12 hours and the absence of a hotel break. Duties included driving, deadheading, rest and idle time subject to constraints on trip coverage, an upper limit on the number of crew members at each base, consecutive working hours, total daily on duty time, and minimum rest time. A column generation-based optimization software package was applied and could solve instances with up to 36,000 trips and 890 stations to 1\% optimality gap within 10 hours. More recently, Jütte and Thonemann (2012) investigated a problem similar to that of Jütte et al. (2011), and developed a divide-and-price algorithm as well as a column generation-based decomposition algorithm. Solutions with a 2\% average optimality gap were obtained within 9 hours. In both cases, the derived schedules were for individual crew districts rather than for the network as a whole, and were of comparable size to the instances we solved.

\indent Shen et al. (2013) conducted research on crew shift scheduling for public transport in China. The transport mode included both passenger trains and buses. The goal was to minimize the number of shifts, total costs of shifts and undercoverage, while satisfying constraints such as the maximum number shifts and minimum demand coverage. An adaptive evolutionary approach incorporating a hybrid genetic algorithm (GA) was developed to solve the problem. In computational experiments with	up to 137355 potential shifts, the proposed algorithm could find better solutions for all the testing problems, compared with the benchmark method of fuzzy GA. Furthermore, the proposed algorithm achieved results close to the lower bounds obtained by a standard linear programming solver in terms of the number of shifts.

\indent The study by Hanafi and Kozan (2014) was based on freight operations in Australia with a one-day planning horizon. A hybrid constructive heuristic coupled with a simulated annealing search algorithm was proposed to make daily trip-duty assignments with the objective of minimizing the total number of duties. Engineers left their home base, completed their duties and returned home. All trips had to be covered by exactly one engineer and deadheads were not allowed. Upper and lower bounds were placed on each duty duration, with the total on duty time similarly limited. For random instances with more than 700 trips, the proposed method could decrease the average excess cost by 3.35\% compared to the manually generated schedules used in practice.

\indent Janacek et al. (2017) developed a step-by-step column generation approach to construct crew schedules for European railway operators. The goal was to minimize total cost of overhead transits while keeping demand covered. In the case study with 75 trains, 2 terminals and 1 week, it was verified that the proposed approach was well-suited for real-life applications.

\indent In a more recent study, Jütte et al. (2017) conducted related research that considered fairness preferences when constructing weekly crew schedules. The objective was to minimize duty durations, number of deadheads, hotel accommodation costs, depot capacity violations, unfairness in the assignment of popular duties, and total unpopularity by selecting low-cost duties that consisted of driving, deadheading and rest. Hard constraints included trip coverage, an upper limit on daily working time, consecutive working time and popular duty time, minimum rest time, and base capacity. The data sets were  provided by European railway freight carriers. A column generation-based algorithm solved instances to within 1.9\% of optimality, on average.

\indent Boyer et al. (2018) investigated a daily vehicle and crew scheduling problem for flexible bus transport system in Mexico. The goal was to assign duties, overtime hours and breaks to drivers, and to minimize the cost of drivers and vehicles. Constraints involved trip coverage, skill qualification and maximum consecutive working time. The authors proposed a mixed-integer linear programming model and a variable neighborhood search (VNS) algorithm to address the problem. In computational experiments, the instances included up to 25 lines and 46 trips per line. Results verified that the mixed-integer linear programming model could not solve instances with more than 2 lines, while VNS could tackle the assignment of 25 lines within 1200 seconds. More recently, Andrade-Michel et al. (2021) developed an exact constraint programming model to construct daily vehicle and reliable drive schedules for public bus transport systems in Mexico. The decision variables and objective function were similar to those of Boyer et al. (2018), while this study also minimized undercoverage. Constraints covered duty length, the maximum consecutive working time, and trip-driver, trip-vehicle and driver-vehicle compatibility. The instances for computations were same as those in Boyer et al. (2018). Compared with VNS, the proposed method could reduce the objective function value by at least 54\% and up to 66\%.

\indent The study by Fuentes et al. (2019) constructed weekly and monthly schedules for passenger train engineers driving between two stations in Spain. The objective was to minimize the number of overnight rests, extra working hours, traveling time to/from the base, the number of deadhead trips and demand undercoverage. Constraints included task sequence, break requirement, maximum duty time, maximum continuous driving time and skill qualification. The authors developed an ad-hoc mathematical decomposition algorithm based on time-personnel clustering, as well as a Fix and Relax Matheuristic (FRA). In the computational experiment, FRA could find solutions with optimality gap of 0.45\% to 70-driver and 1-day instances, while the benchmark algorithm (Branch and Bound) could find solutions with optimality gap of 50.34\%.

\indent Rählmann and Thonemann (2020) investigated a weekly crew scheduling problem for German railway freight carriers considering semi-flexible timetables and time windows that allowed start times to shift forwards and backwards. The objective was to minimize driver wages, fares for deadheading, and artificial costs associated with undesired duties such as those requiring multiple train changes. Other constraints were similar to those of Jütte et al. (2011). The authors used a column generation heuristic to solve instances with up to 1281 trips. The results showed that total cost could be reduced by 4.5\%–9.0\%, compared to the best solutions previously obtained when time shifts were not allowed.

\indent In a more recent study by Rählmann et al. (2021), a column generation method was developed to solve a crew scheduling problem for European railway freight carriers. The objective was to minimize operating cost and the number of overtime hours for a one-week planning horizon. In addition to regular constraints such as mandatory trip coverage, depot capacity, and an upper limit on the number of overtime hours, uncertainty was also taken into account. The computational results showed that for instances with up to 529 trips, the proposed algorithm could improve the objective function by more than 4.2\%, compared with a traditional approach that did not consider uncertainty.

\indent The study by Feng et al. (2023) focused on a crew scheduling and crew rostering problem in urban rail transit of China. The objective was to minimize weighted sum of total travel cost and penalties associated with imbalances in the workloads of crew members. Constraints included maximum accumulated working time and continuous working time, as well as the range of rest time and meal break. An Alternating Direction Method of Multipliers (ADMM)-based dual decomposition mechanism was developed. In computational experiments, the proposed approach obtained an average optimality gap of 4.2\%. This was substantially better than Lagrangian Relaxation, which provided an average optimality gap of 34.73\%.

\indent Gattermann-Itschert et al. (2023) investigated a European freight crew scheduling problem for a one-day planning horizon with the objective of minimizing operating cost (including wages and hotel stays) and preference violations (such as long duties). Constraints included depot capacity, upper limit on the number of working hours, and mandatory trip coverage. The authors proposed an algorithm that combined machine learning and optimization to solve instances with more than 200 trips. The computational results showed that the solution for the acceptance probability, which was linked to preference violations, was improved by 12\% with little increase in operating cost when compared to current practice.

\indent Feng et al. (2024) proposed a branch-and-price algorithm to construct 4-day crew schedules for China urban railroad. The objective was to minimize the total operational and labor costs. Constraints included trip coverage, upper limit on the number of engineers, accumulated working time, continuous working time, and fixed ranges of rest time and lunch time. In the computational experiments, the branch-and-price algorithm was able to solve problems with up to 248 engineers. The average optimality gap was 3.28\%. Compared with Lagragian Relaxation and Alternating Direction Method of Multipliers, the proposed algorithm performed 16.4\% and 5.03\% better, respectively. 

\indent In the study by Guo and Bard (2024), a 3-phase algorithm that relied on the logic of column generation and local improvement procedures was developed to construct weekly schedules and allocate home cities for freight engineers in a U.S. Class I railroad. Their multi-objective formulation was designed to minimize the number of engineers traveling between two cities and the number of deadhead trips, and to maximize demand coverage and the number of times engineers with consecutive 48-hour rest. Hard constraints consisted of the fraction of engineers assigned to each city, a lower limit on the number of trips, a minimum rest time between two trips, and an upper limit on the number of consecutive driving hours, the number of driving hours in a week and the number of rest hours in the away base. In computational experiments with up to 207 trips, the proposed algorithm was able to construct weekly schedules with trip coverage rate between 95.29\% and 99.52\%. With respect to minimizing the number of required engineers, the optimality gap was less than 4\%, which resulted in a much smaller number of engineers than was employed by the company who supplied the data. The study was intended as a ``proof of concept'' to show that it is possible to construct traditional weekly schedules for freight engineers when demand is relatively firm. This is rarely the case in the U.S., though, so railroads have yet to implement 5-day work weeks for their drivers.

\indent Wang et al. (2024) developed a branch-price-and-cut method and a heuristic to solve a robust safety driver scheduling problem for autonomous buses in Singapore. The goal was to make task-duty assignment, and to insert breaks between tasks in a duty. The obejctive was to minimize the total delay and duty cost while satisfying constraints such as the maximum continuous workload for each driver, maximum number of safety drivers and break requirement. Instances in computation experiments were from Singapore's bus line with up to 350 tasks. Results showed that for large-scale instances, the proposed heuristic method outperformed the benchmark method (Shortest-path and Matching) in terms of solution quality. However, this advantage came with the price of slightly longer computational times. 

\indent In the study by Xu et al. (2025), a passenger rail crew scheduling problem with duty and break assignment was investigated. The objective was to minimize the number of crew members and operating cost, improve workload balance, and reduce waiting time. Crew members were allowed to operate across multiple lines to improve efficiency. Constraints included trip coverage, maximum working time per duty, limits on continuous work, and minimum break time. An extended set partitioning model and a column generation algorithm were developed to solve the problem. In computational experiments, compared with traditional single-line scheduling, the proposed algorithm significantly reduced the number of crew members, lowered operating cost, and improved workload balance.

\indent More recently, Lyu and Bard (2025) investigated a related freight rail crew scheduling problem based on the work of Guo and Bard (2024). The objective function, decision variables and constraints were similar to those presented in the former study but the authors extended the problem definition to include three cities rather than just two. A network approach was developed to construct weekly schedules and assign home base to engineers. In computational experiments with up to 207 trips for 2-city problems and 323 trips for 3-city problems, the results verified that optimal weekly schedules could be constructed in minutes for engineers in crew districts with two cities, and in several hours for engineers in crew districts with three cities.

\subsection{Personnel Scheduling with Call Windows in Freight Rail and Other Fields} 
\label{sec_literature_window}

\noindent In this subsection, we concentrate on personnel scheduling with call windows that do not include detailed duty assignments within the predetermined interval. After careful review of the existing literature, we have found only a handful of studies on call window scheduling and only one related to freight rail, underscoring a significant gap in the literature. The discussion is in chronological order.

\indent Shamia et al. (2015) investigated a call window scheduling problem in healthcare. The authors assigned on call schedules to physicians with different skills and seniority levels. The goal was to generate on call schedules that satisfied as many individual preferences and duty requirements as possible, while ensuring optimum usage of available resources to guarantee a better service for the patients. The constraints included limits on the number of consecutive working hours, the number of shifts for each physician, number of shifts assigned, vacation periods, weekend off requests and balanced work load. Mathematical programming models were developed to provide solutions. In the computational experiments, the data set was a real case from a Qatari hospital with 2 skills, 2 seniority levels, 16 physicians and a 2-month planning horizon. Compared with the manual approach in use at the time, the proposed algorithm could significantly reduce the time and effort required to construct schedules that met both physician and hospital requirements.

\indent In the study by El-Rifai et al. (2016), a two-stage stochastic integer linear program and a sample average approximation method were proposed to assign regular and on call duties to employees in healthcare. The objective was to minimize the number of physicians over the planning horizon while trying to meet as much demand as possible. Constraints included the minimum number of regular on-duty periods, an upper limit on the number of on call duties and night shifts for each employee, incompatibility between certain duties, and a minimum rest time. In the computational experiments, data were provided by an emergency department in Lille, France. The results verified that when resource shortages were very expensive, using on call duties could lead to more than a 30\% reduction in effective work hours with schedules that were 10\% less expensive.

\indent Scherer et al. (2021) developed an on call scheduling procedure on ground support for space operations. On call shifts were assigned for full days, where each operator could cover a certain subset of positions depending on their training and skill set. The objective was to have operators work the same number of days each week. Constraints involved position requirement, demand coverage, and a maximum number of working days in certain periods. The authors developed Grover's algorithm (see Grover 1996) to solve the optimization problem. Computational experiments were implemented using data set from the German Space Operation Center with up to 8 operators, 3 positions and 6 days. The proposed algorithm achieved success rates between 88\% and 99\% for valid configurations, where the ``success rate" is calculated by running the algorithm 8000 times and determining the relative number of valid schedules.

\indent The study by Frisch et al. (2022) centered on a weekly crew scheduling problem that assigned shifts to freight engineers for Rail Cargoo Austria (RCA). The objective was to minimize overall paid working time. Constraints included a fixed number of unpaid breaks as well as an upper and lower limit on shift lengths. The authors developed a set partitioning model for the problem and used a breadth-first search construction heuristic to find solutions. The computational experiments verified that the proposed approach was suitable for a real-life application for RCA as it delivered good solutions within reasonable time for instances with up to 172 trips. 

\indent Van Rossum et al. (2025) investigated a template (call window) scheduling for passenger rail crews in the Netherlands. The goal was to assign templates and duties to crew members so that the template costs and recovery costs could be minimized. Each duty must start and end at the same base, and it should be covered by either available templates or excess duties. Constraints included upper limit on the number of template types, duty length and recovery cost. A 30-minute meal break was also considered in each duty. The authors developed a two-phase accelerated Benders decomposition algorithm. In computational experiments, the instances had up to	948 tasks per day. Compared with a literature benchmark, the proposed method solved three times as many instances without rostering constraints to optimality.

\subsection{Research Gap} \label{sec_literature_gap}

\noindent Our study contributes to the state of the art in freight rail crew scheduling by developing a new approach to workforce and home base assignments that is based on call windows rather than specific trips. Despite a significant amount of prior work on freight rail crew scheduling, only the paper by Frisch et al. (2022) made use of call windows, but then only for limited work rules and without considering home base assignments..

\indent In this study, our goal is to optimize call window and home base assignments, while taking into account critical personnel factors such as the maximum number of consecutive on call days, same on call pattern for each day, demand coverage, number of engineers, preparation time, and minimum rest time between call windows. It should be noted that in this study, we only optimize the call window schedules and home base of engineers, and do not construct the detailed duty within each call window. For example, we neither make trip-engineer assignments nor decide the rest time, rest location and deadhead schedules. However, in the experiment component of the computational experiments, we consider all the duty details that are discussed in the paper of Guo and Bard (2024). Table \ref{tbl_literature_oncall} summarizes the similarities and differences between this paper and prior work for call window scheduling without consideration of detailed duty optimization within the call windows.

\begin{table}[htbp]
	\centering
	\caption{Summary of differences and similarities between this study and prior work for call window scheduling without consideration of detailed duty optimization within call windows} 
	\resizebox{\textwidth}{!}{%
		\begin{tabular}{|p{4cm}|p{6cm}|p{2cm}|p{2cm}|p{2cm}|p{2cm}|p{2cm}|p{2cm}|}
			\hline
			\multirow{2}{*}{Problem components} & Features & This work & \multicolumn{1}{p{4.215em}|}{Shamia \newline{} et al. \newline{}(2015)} & \multicolumn{1}{p{4.215em}|}{El-Rifai \newline{} et al. \newline{}(2016)} & \multicolumn{1}{p{4.215em}|}{Scherer \newline{} et al. \newline{}(2021)} & \multicolumn{1}{p{4.215em}|}{Frisch \newline{} et al. \newline{}(2022)} & \multicolumn{1}{p{4.215em}|}{van Rossum \newline{} et al. \newline{}(2025)}\\
			\cline{2-8}  & Industry & Freight rail & Healthcare  & Healthcare  &  Spacecraft & Freight rail & Passenger rail \\
			\hline \multirow{2}{*}{Variables} & Call window assignment & \checkmark  &  \checkmark &  \checkmark &  \checkmark &  \checkmark & \checkmark \\
			& Home base assignment & \checkmark  &   &   &   &   &  \\
			\hline \multirow{5}{*}{Constraints} & Demand coverage  & \checkmark  &   &   &  \checkmark &  &  \checkmark \\
			& Same on call pattern  & \checkmark  &   &   &   &   &  \\
			& Consecutive on call time  &  \checkmark & \checkmark  & \checkmark  & \checkmark  & \checkmark   & \\
			& Total on call time  & \checkmark  & \checkmark  & \checkmark  &   &   &  \\
			& Rest time between call windows  &  \checkmark &   & \checkmark  &   &   &  \\
			\hline \multirow{3}{*}{Objective} & Max. demand coverage & \checkmark  &   & \checkmark  &   &   & \checkmark \\
			& Min. no. employees &  \checkmark &   & \checkmark  &   &   &  \checkmark \\
			& Other penalties &   & \checkmark  &   & \checkmark  & \checkmark   &  \\
			\hline
		\end{tabular}
	}
	\label{tbl_literature_oncall}%
\end{table}%

\section{Problem Description} \label{sec_probdes}

\noindent A trip is defined by its origin, destination, departure and arrival times. The arrival times are assumed to be uncertain due to sporadic congestion and multiple disruptions during the day. The objective is to find a balance between trip coverage and labor cost while keeping all hard constraints satisfied. At each station or terminal in a company's network, upcoming trips and available drivers are displayed on an electronic ``Board'' or queue that is maintained by the operations manager. Driver availability is determined by the HOS regulations.  Dispatchers in the operations control center alert the local drivers of their next assignment either by texting or calling them.

\indent As the rail industry is currently structured, each engineer works in a district consisting of two or three cities, and is assigned one of those cities as his home base. The other cities are termed ``away bases.'' We maintain this structure but instead of requiring each driver to be on call 24/7 (essentially, a 24-hour call window) subject to the HOS, our goal is to construct a call window schedule for all engineers that guarantees at least two days off in every week.  Engineers at their home base are expected to be available 3 hours after the start of their call window. Assignments are based on a first in, first out rule; that is, the driver on the Board the longest is assigned the next trip if its departure time is within his call window plus 3 hours.  If not, he is removed from the Board (queue) until his next call window opens up.  For trips from the home base  to an away base, the driver can either deadhead by taking a van home, or rest at least 10 hours and then drive a train back home if timing permits.

Note that we only allow deadheading by van. Deadheading by train is not considered due to a combination of efficiency, reliability, and cost. Specifically, van transport offers significant time savings, as trains often follow longer, indirect routes compared to the point-to-point efficiency of highways. This is compounded by the ever-present issue of freight rail delays, which make train-based deadheading highly unpredictable and disrupt carefully planned engineer schedules. This can lead to costly overtime or crew shortages. The economic rationale is strengthened by volume-based contracts with transport companies, making the van option more attractive.

\indent In this study, we investigate two cases: (i) districts with 2 cities  (cities A and B), and (ii) districts with 3 cities (cities A, B and C). In case (ii), city B connects cities A and C, i.e., there are trips between cities A and B, and between cities B and C, but no trips between cities A and C. If an engineer has his home base in city B, then he can drive  between A and B, and B and C; however, if an engineer's home base is in city A, then he cannot drive trips between B and C. Similarly, an engineer residing in city C cannot drive between A and B. Moreover, in both cases (i) and (ii), an engineer cannot take a van from his home to away base. In the remainder of this section, we will introduce decisions variables, constraints, objective function and assumptions.

\subsection{Decision Variables} 
\noindent For each engineer, we need to make two sets of decisions: (1) home base assignment: each engineer must be assigned exactly one home base (City A, B or C); and (2) call window schedules: the days and starting time of call windows. 

\subsection{Constraints} \label{sec_probdes_cons}

\noindent For each engineer, the call window schedule must satisfy the following hard constraints.\\[-20pt]
\begin{enumerate}
	\item An engineer must be assigned to exactly one home base. \\[-20pt]
	\item An engineer cannot be on call for more than 5 days in a row. . \label{conswindow_consecutive} \\[-20pt]
	\item In each week, there must be at least two days on which no call window starts.\label{conswindow_minoff} \\[-20pt]
	\item For any call window schedule, the call window is the same on each working day unless the day is off. Accordingly, each engineer has the same call window on every working day, except on days off. \label{conswindow_same} \\[-20pt]
	\item Among all the engineers, the call window length (a parameter) is fixed. For example, if the call window is 10 hours, then the call window length for all engineers on their on call days must be exactly 10 hours.  \\[-20pt]
\end{enumerate}

\subsection{Objective}

\noindent The objective function is to minimize the weighted sum of two terms: (1) the number of uncovered trips and (2) the number of engineers who are on call in the planning horizon (size of workforce). 

\subsection{Assumptions} \label{sec_probdes_assum}
\begin{enumerate}
	\item If a trip is not covered in the final set of schedules, it will be covered by an ``extra Board'' or ``eBoard'' engineer. This is what is done now. An eBoard engineer has other duties in the railyard but is available to drive on short notice.  \\ [-20pt]
	\item The call window schedules are acyclic, i.e., they do not repeat every 14 days.  \label{assumptionwindow_cyclic} \\[-20pt]
	\item Each day starts at 7:00 am.  \\[-20pt]
	\item The planning horizon is 2 weeks (14 days).   \\[-20pt]
	\item During the call window, an engineer in his home base is on call after resting for at least 10 hours but is only eligible for his next trip after 3 additional hours (allowed for preparation). \\[-20pt]
	\item During the call window, an engineer in his away base is on call after resting for 10 hours but is only eligible for his next trip after 1.5 additional hours. \\[-20pt]
	\item Regarding an engineer's  status, only driving and taking a van are considered ``working." Being on call while not driving/taking a van is not ``working." \\[-20pt]
	\item Assumptions for engineer-trip assignment \\
	 In the computational experiments, the following conditions must be satisfied. \\[-20pt]
		\begin{enumerate}
		\item If an engineer drives  from his home to an away base, and it is decided that he returns home by van, then the van trip starts when the driven train arrives at the away base. Taking a van from the home to away base is not permitted.
		\item Hours of Service (HOS) Regulations. 
		\begin{enumerate}
			\item Driving restrictions \\ [-20pt]
			\begin{enumerate}
				\item An engineer cannot drive more than 12 consecutive hours.
				\item An engineer cannot drive more than 70 hours in a week.
			\end{enumerate}
			\item Deadhead restrictions \\
			During each week, an engineer cannot spend more than 7.5  hours waiting for or in deadhead transportation from a duty assignment to the place of final release following a period of 12 consecutive hours on duty. For example, assume an engineer has the following partial duty. \\ [-15pt]
			\begin{enumerate}
				\item[$\bullet$] drives a train that departs his home base at  8:00 am on Monday,
				\item[$\bullet$] arrives at his away base at 8:00 pm on Monday,
				\item[$\bullet$] takes a van departing the away base at  8:00 pm on Monday, and
				\item[$\bullet$] arrives at his home base on Monday at 11:00 pm.
			\end{enumerate}
			In this example, 15 hours transpires from the time the engineer departs his home base on Monday 8:00 am until he returns on Monday 11:00 pm. Therefore, $15-12=3$ hours will be counted towards the 7.5-hour limit given that the ``12 consecutive hours on duty" extends from 8:00 am to 8:00 pm on Monday. Any consecutive on-duty hours after  8:00 pm Monday is counted towards the 7.5-hour limit.
			\item Rest rules  \\ [-20pt]
			\begin{enumerate}
				\item An engineer must have at least 10-hour rest between every two driven trips.
				\item An engineer cannot rest for more than 15 hours in a row at an away base.
				\item When an engineer has worked 6 consecutive days, he must rest for at least 48 consecutive hours after the $6^{\text{th}}$ working day. When an engineer has worked 7 consecutive days, he must rest for at least 72 consecutive hours after the $7^{\text{th}}$ working day.  Although these constraints are part of the HOS, they do not apply to our situation because we allows or at most 5 working days in a row.
			\end{enumerate}
		\end{enumerate}
	\end{enumerate}
\end{enumerate}

	Note, that the trip-engineer assignment assumptions are not considered when we construct call window schedules because they are applied only when an engineer is performing specific tasks (e.g., drive, deadhead, and rest) during his call window.  As part of  the computational experiments, simulation is performed to validate the results obtained from our optimization models and to address demand uncertainty.\\[-20pt]

\subsection{Illustrative Example}

\noindent This subsection presents a small example to illustrate the schedule and call windows that satisfy the problem constraints and assumptions.

\indent Assume we have 4 engineers: $E1$, $E2$, $E3$ and $E4$, and there are 2 cities in the system: $A$ and $B$. Engineers $E1$ and $E2$ have their home base in city $A$, while engineers $E3$ and $E4$ have their home base in city $B$. For simplicity, we only consider 3 days in the planning horizon: day 1, day 2 and day 3. The intervals for days 1, 2 and 3 are (Mon. 7am, Tue. 7am), (Tue. 7am, Wed. 7am) and (Wed. 7am, Thu. 7am), respectively. Assume all call window lengths are 10 hours and that there is no minimum number of required working days. The demand consists of 6 trips ($T1$, ..., $T6$) specified as follows. \\[-20pt]

\begin{enumerate}
	\item Trip $T1$ starts on Wednesday 5:12 am in city $A$ and ends on Wednesday 11:47 am in city $B$. \\[-20pt]
	\item Trip $T2$ starts on Tuesday 1:21 pm in city $A$ and ends on Tuesday 8:48 pm in city $B$. \\[-20pt]
	\item Trip $T3$ starts on Wednesday 9:47 am in city $B$ and ends on Wednesday 4:35 pm in city $A$. \\[-20pt]
	\item Trip $T4$ starts on Tuesday 1:41 pm in city $B$ and ends at Wednesday 12:36 am in city $A$. \\[-20pt]
	\item Trip $T5$ starts on Tuesday 2:49 am in city $B$ and ends at Tuesday 9:11 am in city $A$. \\[-20pt]
	\item Trip $T6$ starts on Wednesday 4:44 am in city $B$ and ends at Wednesday 3:39 pm in city $A$. \\[-10pt]
\end{enumerate}

\begin{figure}[!h]
	\centering
	\includegraphics[width=7.5in]{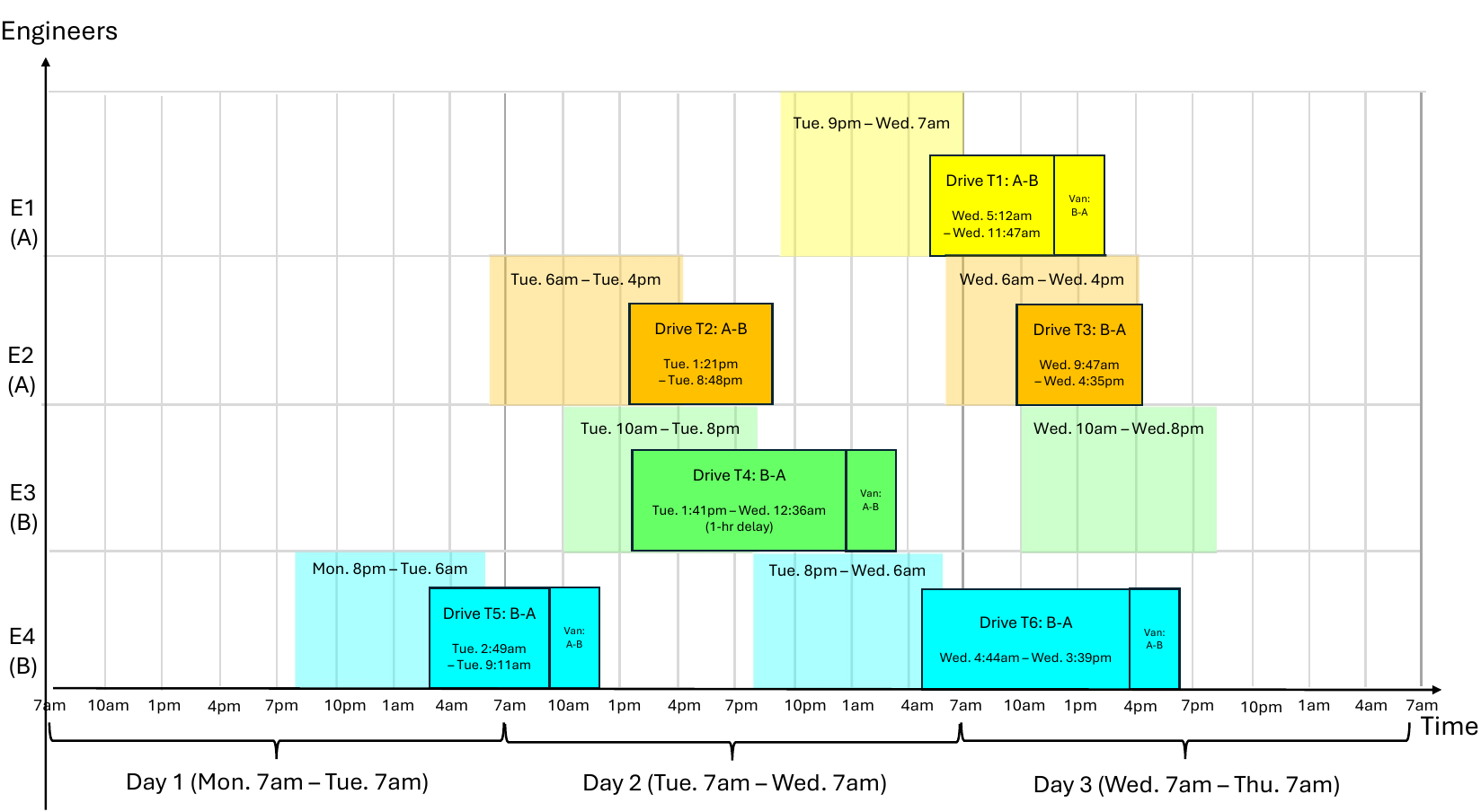}
	\caption{Illustrative example of a 2-city instance.}
	\label{Fig_schedule_2city}
\end{figure}

\indent In Figure \ref{Fig_schedule_2city}, the horizontal and vertical axis present the time and engineers, respectively. Transparent colored boxes indicate the call windows of the engineers, and the nontransparent colored boxes give details of trips assigned to the corresponding engineer, including origin, destination, start time, end time and delay information.

\indent As we can observe from the figure, one feasible call window schedule for the 4 engineers is: engineer $E1$ is on call from Tuesday 9:00 pm to Wednesday 7:00 am; engineer $E2$ is on call from 6:00 am to 4:00 pm on Tuesday and Wednesday; engineer $E3$ is on call from 10:00 am to 8:00 pm on Tuesday and Wednesday; and engineer $E4$ has a 10-hour call window that starts at 8:00 pm on both Monday and Tuesday.

\indent Based on the call windows, trip $T1$ is assigned to engineer $E1$, and $E1$ takes a van back to city $A$ after the trip. Engineer $E2$ drives trip $T2$ from her home base (city $A$) during her call window on Tuesday. After she arrives in city $B$ at 8:48 pm on Tuesday, she rests for 12 hours 59 minutes and drives trip $T3$ back to her home base. Trip $T4$ is assigned to engineer $E3$, who takes a van home afterwards. Note that $T4$ is delayed by 1 hour because the call window of engineer $E3$ starts at 10:00 am and she cannot drive from her home base unless she is notified at least 3 hours in advance. The other point we would like to highlight is that even though engineer $E3$ has a call window between 10:00 am and 8:00 pm on Wednesday, she is not assigned a trip during that interval.  Finally, engineer $E4$ drives trip $T5$ during the Monday call window, takes a van home, drives trip $T6$ during the Tuesday call window, and again takes a van home .

\section{Methodology 1: Set-covering-type Model} \label{sec_method}

\noindent The basic idea of the methodology is to first construct all feasible call window schedules that satisfy the constraints in Section \ref{sec_probdes_cons}, then, for engineers at each base, use a set-covering-type model to select the ``best" schedules that balance trip coverage and labor cost.

\indent The following notation is used in the developments. \\
\\
\begin{minipage}{\columnwidth}
	\noindent \textit{Sets and indices} \\
	\setlength{\parindent}{0em}
	\indent
	\begin{tabularx}{\textwidth}{p{2.2 cm}X}
		$d$		& index for day \\
		$D$		& set of days in the planning horizon \\
		$d^{\text{first}}, d^{\text{last}}$ & indices for the first and last day in the planning horizon. $d^{\text{last}} - d^{\text{first}} = |D|-1$. \\
		$D^{\text{I}}$, $D^{\text{II}}$ & set of first and second half days in the 14-day planning horizon \\
		$d_{1,1}, \; d_{1,2} $		& indices for any two days in $D^{\text{I}}$ \\
		$d_{2,1}, \; d_{2,2} $		& indices for any two days in $D^{\text{II}}$ \\	
		$h$		& index for hour \\
		$H$		& set of hours in the planning horizon \\
		$i$     & index for engineer \\
		$I_b$	    & set of engineers whose home base is city $b$ \\
		$s$		& index for a call window schedule (set of days and eariest start times) \\
		$S$		& set of call window schedules \\
	\end{tabularx}
\end{minipage}
\\
\\
\begin{minipage}{\columnwidth}
	\noindent \textit{Parameters} \\
	\setlength{\parindent}{0em}
	\indent
	\begin{tabularx}{\textwidth}{p{2.2 cm}X}
		$a_s$ & consecutive number of working days in call window schedule $s$ at the end of the previous planning horizon. \\
		$\alpha^{\text{U}}$  & penalty weight for each uncovered trip \\
		$c_d$  &  1 if the engineer has call window on day $d$, 0 otherwise \\
		$c$  &  vector consisting of components $c_d \; (d \in D)$ that reflects whether or not an engineer has a call window on each day of planning horizon; for example, $c=(1,1,0,1,1,1,0,1,1,1,1,0,0,1)$ indicates that days 1, 2, 4-6, 8-11, 14 are assigned call windows \\
		$C$ & set of feasible vectors that reflect whether the engineer has a call window on each day \\ 
		$\bar{d}$  & maximum number of days an engineer can be on call in a row; in our problem, we have $\bar{d}=5$  \\
	\end{tabularx}
\end{minipage}
\begin{minipage}{\columnwidth}
	\setlength{\parindent}{0em}
	\indent
	\begin{tabularx}{\textwidth}{p{2.2 cm}X}
		$\delta_{\text{H}}$ &  minimum number of hours an engineer can be on the Board in his home base before he can begin  a   trip; in our problem, $\delta_{\text{H}} = 3$ \\
		$k_{sh}$		& 1 if engineers with call window schedule $s$ can drive trips starting in hour $h$, 0 otherwise  \\
		$\bar{l}^{\text{trip}}$ & average trip length \\
		$l^{\text{window}}$  & length of a call window; in the study, we perform a parametric analysis for  $l^{\text{window}}=$  6, 10, 12 and 24 hours \\
		$s^{\text{day}}$ & start time of each day; in our problem, we set $s^{\text{day}} = 7$. \\
		$w_{bh}$		& number of trips starting in hour $h$ from base $b$ \\ [8 pt]
	\end{tabularx}
\end{minipage}
\\
\\
\begin{minipage}{\columnwidth}
	\noindent \textit{Decision variables} \\
	\setlength{\parindent}{0em}
	\indent
	\begin{tabularx}{\textwidth}{p{2.2 cm}X}
		$U_h$		& (nonnegative integers) the number of uncovered trips starting in hour $h$  \\
		$X_s$		& (nonnegative integers) the number of engineers assigned to call window schedule $s$  \\
	\end{tabularx}
\end{minipage}

\subsection{Construction of feasible call window schedules}

\noindent The construction of a feasible call window schedule consists of two steps: (1) select four days off in the planning horizon that satisfy Constraints \ref{conswindow_consecutive} and \ref{conswindow_same} in Section \ref{sec_probdes_cons}, and Assumption \ref{assumptionwindow_cyclic} in Section \ref{sec_probdes_assum}; (2) select $l^{\text{window}}$ consecutive hours starting on each working day that satisfy Constraints \ref{conswindow_consecutive} and \ref{conswindow_same}, Assumption \ref{assumptionwindow_cyclic}, and working days selected in Step \ref{stepwindow_day}. Further explanation of these steps is given below; the pseudocode for the construction is given in  Algorithm \ref{alg_window}.

\begin{enumerate}
	\item Select days. \label{stepwindow_day}\\
	Constraint \ref{conswindow_consecutive} and Assumption \ref{assumptionwindow_cyclic} imply that the first $\frac{|D|}{2}$ days and the second $\frac{|D|}{2}$ days in the planning horizon must have 2 days off, respectively. Therefore, the number of potential schedules is ${\frac{|D|}{2} \choose 2} \times {\frac{|D|}{2} \choose 2}$. Since the planning horizon has 14 days ($|D|=14$), we have 441 potential schedules. After enumerating all such schedules, we check whether each  satisfies Constraint \ref{conswindow_consecutive} and Assumption \ref{assumptionwindow_cyclic}. Let $c_d$ be 1 if $d$ is a working day, 0 otherwise. Only when schedule $s$ satisfies the following inequalities does it meet the requirements in Constraint \ref{conswindow_consecutive} and Assumption \ref{assumptionwindow_cyclic}.
	\begin{subequations} \label{windowday_constraints}
		\begin{alignat}{3}
		& \sum\limits_{d'=d}^{d+\bar{d}} c_{d'} \leq \bar{d} & \qquad \qquad  \forall \; d = d^{\text{first}},  d^{\text{first}}+1, ..., d^{\text{last}}-\bar{d} \label{windowday_constraints_a} \\
		& \sum\limits_{d'=d^{\text{first}}}^{d^{\text{first}}+\bar{d}-a_s} c_{d'} \leq \bar{d} - a_s &  \label{windowday_constraints_b}
		\end{alignat}
	\end{subequations}
	\\
	If a schedule does not satisfy Eqs. \eqref{windowday_constraints_a} and \eqref{windowday_constraints_b}, then it is not feasible  and should be removed from the set of candidates. Now, as previously defined, let $c$ be a binary vector of size $|D|$ such that $c = (c_{d^\text{first}}, …, c_{d^\text{last}})$.

	\item Select hours. \\
	We use Step 2 in Algorithm \ref{alg_window} to select call windows, based on the set of candidate combination of days on/off obtained in Step \ref{stepwindow_day}.
\end{enumerate}

\begin{algorithm}
	\begin{algorithmic}[1]
		\caption{Call window construction}
		\label{alg_window}
		\State \textbf{Step 1. Select Days}
		\State $C \leftarrow \phi$
		\State Let $D^{\text{I}} = [d^{\text{first}},...,d^{\text{first}}+\frac{|D|}{2}]$, $D^{\text{II}} = [\frac{|D|}{2}+1,...,d^{\text{last}}]$ \vspace{0.5em}
		\State (Split the planning horizon into two halves: $D^{\text{I}}$ and $D^{\text{II}}$). \vspace{0.5em}
		\For{$(d_{1,1}, d_{1,2})_{d_{1,1} \neq d_{1,2}} \in D^{\text{I}}$, $(d_{2,1}, d_{2,2})_{d_{2,1} \neq d_{2,2}} \in D^{\text{II}}$ \vspace{0.5em}  \\ 
		\qquad (for any 2 different days in $D^{\text{I}}$ and $D^{\text{II}}$, respectively)}
			\State $c_{d_{1,1} \leftarrow 0}$, $c_{d_{1,2} \leftarrow 0}$, $c_{d_{2,1} \leftarrow 0}$, $c_{d_{2,2} \leftarrow 0}$ 
			\State (let the engineer rest on these 4 days)
			\For{$d \in D \backslash \{d_{1,1}, d_{1,2}, d_{2,1}, d_{2,2}\}$ }
				\State $c_d \leftarrow 1$ 
				\State (assign call windows to the engineer on the other days)
			\EndFor
			$c \leftarrow (c_d)_{d = d^{\text{first}}, ..., d^{\text{last}}}$    \\
			\quad \; (construct vector $c$ that reflects whether the engineer has call window on each day)
			\If{$c$ satisfies constraints \eqref{windowday_constraints_a} and \eqref{windowday_constraints_b}}
				$C \leftarrow C \cup \{c\}$ \\
				\qquad \; (if $c$ is a feasible vector of call window days, then add it to $C$)
			\EndIf
		\EndFor
		
		\\
		\State \textbf{Step 2. Select Hours}
		\State $S \leftarrow \phi$
		\For{$c \in C$}
			\For{$\Delta \in [0,1,2,...,23]$}
				\For{$h \in H$}
					$k_h \leftarrow 0$
				\EndFor
				\For{$d \in D$}
					\If{$c_d=1$}
						\If{(1) the end of the potential time window is on day $d$ ($s^{\text{day}}+\Delta + l^{\text{window}} \leq 24 + s^{\text{day}} $) \\
							\qquad \qquad \qquad \; \; or (2) day ($d+1$) is selected a working day ($c_{d+1}=1$)}
							\State Assign a call window starting at time $s^{\text{day}}+\Delta$ on day $d$
							\State Let $\hat{H}$ be hours covered by the call window starting at  time $s^{\text{day}}+\Delta$ on day $d$
							\State Let $\hat{H}_{\delta}$ be hours with $\delta_{\text{H}}$-hour lag to $\hat{H}$
							\For{$h \in \hat{H}_{\delta}$}
								$k_h \leftarrow 1$
							\EndFor
						\EndIf
					\EndIf
				\EndFor
				\State Let $s$ be the constructed call window schedule
				\For{$h \in H$}
					$k_{sh} \leftarrow k_h$
				\EndFor
			\EndFor
		\EndFor
	\end{algorithmic}
\end{algorithm}

\subsection{Set-covering-type Model}

\noindent For engineers whose home base is  $b$, Model (\eqref{windowmodel}) selects the ``best" call window schedules that balance trip coverage and labor cost. The number of selected call window schedules is the number of on call engineers, which is a variable with an upper bound of $|I_b|$.\\

\noindent \textit{Call Window Selection Model for base $b$} \\
\begin{subequations} \label{windowmodel}
	\begin{alignat}{3}
	\text{Minimize} \qquad \sum\limits_{s \in S} X_s  +  \alpha ^{\text{U}} \sum\limits_{h \in H} U_h  \label{eq:windowmodel_obj}
	\end{alignat}
	Subject to 
	\begin{alignat}{3}
	& \sum\limits_{s \in S} k_{sh}X_s  + U_h \geq w_{bh}  &    \qquad \qquad \qquad \qquad \forall \; h \in H \label{eq:windowmodel_demand} \\
	& \sum\limits_{s \in S} X_s \leq |I_b| & \label{eq:windowmodel_engineerbound} \\
	& U_h \geq 0, \; X_s = 0, 1, 2, ... & \qquad \qquad \qquad \qquad \forall \; s \in S, \; h \in H \label{eq:windowmodel_vardef}
	\end{alignat}	
\end{subequations}

\indent The objective function \eqref{eq:windowmodel_obj} consists of two terms. The first  reprensents the number of selected call window schedules, which equals the number of engineers who will be on call in the planning horizon. The second corresponds to weighted trip undercoverage, which is the weight ($\alpha^{\text{U}}$) multiplied by the number of uncovered trips. Constraints \eqref{eq:windowmodel_demand} define the number of uncovered trips in hour $h$ by correlating variables $X$ and $U$. Constraints \eqref{eq:windowmodel_engineerbound} indicate that the number of selected call windows (selected engineers) cannot exceed the maximum number of engineers who can be assigned to home  base  $b$. Constraints \eqref{eq:windowmodel_vardef} define variables $U$ and $X$.

\section{Methodology 2: Direct Algorithm} \label{sec_method2}

\noindent In this approach, an iterative algorithm is run to determine the minimum  number of engineers and their call window schedules needed to cover demand. The algorithm starts by assigning a certain number of engineers to each of the cities in the district being modeled, and solves a 0-1 integer program [Model (\ref{constructcwmodel}) given below] to determine their call window schedules. Note that the number of engineers who are assigned to each city is a an input parameter that can be adjusted by the user.
	
After initialization, if there are still uncovered trips and the total number of engineers has not reached its upper limit, then at subsequent iterations, we add a fixed number of engineers (again, a parameter) to each home base and solve Model (\ref{constructcwmodel}) to generate a call window schedule for each engineer. The new engineers will be assigned some of the \textit{remaining uncovered trips}. The trip-engineer assignment procedure is outlined in Algorithm \ref{alg_simulation_itr} in the next section. The steps of the Direct Algorithm are given below; termination occurs when either all trips are covered or the total number of engineers reaches a user defined upper limit. 

	\begin{enumerate}
		\item (Initialization) In each city $b$, let $I_b=\phi$, where $I_b$ is the set of new engineers whose home base is  $b$. Assign a predetermined number of engineers to set $I_b$.  \label{directalg_step_init_engineer} \\[-20pt]
		\item For each base $b$ with the set of new engineers $I_b$,  solve Model (\ref{constructcwmodel}) to determine the call window schedules of engineers in $I_b$. \label{directalg_step_init_model} \\[-20pt]
		\item Assign uncovered trips to engineers in $I_b$ with call window schedules obtained in Step \ref{directalg_step_init_model}. The trip-engineer assignment mechanism is described in lines 11-39 of Algorithm \ref{alg_simulation_itr}. \label{directalg_step_assign} \\[-20pt]
		\item Update the list of uncovered trips. \label{directalg_step_init_update} \\[-20pt]
		\item If there are still uncovered trips and the total number of engineers has not reached the upper limit, add a predetermined number of engineers and go to Step \ref{directalg_step_init_model}; otherwise, stop. \label{directalg_step_init_stop} \\[-20pt]
		
	\end{enumerate}

In this procedure, there are two issues that need further clarification. First, in both Steps \eqref{directalg_step_init_engineer} and \eqref{directalg_step_init_stop}, the number of new engineers added to each base $b$ is a parameter set by the user. In our computations, to avoid unnecessary labor costs, we add only one new engineer to each base per iteration, i.e., $|I_b|=1 \; \forall  \, b \in B$. Second, even though Model (\ref{constructcwmodel}) generates feasible call window schedules for the engineers in $I_b$, the engineer-trip assignment obtained from the model can be infeasible. This is because it does not consider HOS regulations such as a 10-hour rest period between trips. The purpose of Model (\ref{constructcwmodel}) is to provide a feasible call window schedule for each engineer in $I_b$, with the immediate objective of maximizing trip coverage. Hence, it does not guarantee feasibility of the trip-engineer assignment, which is ultimately obtained in the steps in lines 11-39 of Algorithm \ref{alg_simulation_itr}. \\
\indent The following notation is used to present the  call window construction model. \\
\\
\begin{minipage}{\columnwidth}
	\noindent \textit{Sets and indices}  \\
	\setlength{\parindent}{0em}
	\indent
	\begin{tabularx}{\textwidth}{p{2.2 cm}X}
		$FH_n$  & set of first $n$ hours in planning horizon \\
		$H_d$	& set of hours on day $d$ \\
		$I_b$	& set of new engineers whose home base is $b$ \\
		$LH_n$  & set of last $n$ hours in planning horizon \\
		$LD_n$	& set of last $n$ days in planning horizon \\
		$o_t$, $d_t$ & origin and destination of trip $t$ \\
		$t$		& index for trip t \\
		$T$		& set of trips \\
		$T^{\text{U}}_b$		& set of uncovered trips starting from base $b$ \\[6pt]
	\end{tabularx}
\end{minipage}
\\
\begin{minipage}{\columnwidth}
	\noindent \textit{Parameters} \\
	\setlength{\parindent}{0em}
	\indent
	\begin{tabularx}{\textwidth}{p{2.2 cm}X}
		$c^{\text{day}}$  & maximum number of consecutive days an engineer can be given a call window\\
		$\underline{d}^{\text{rest}}$ & minimum number of rest days in the planning horizon \\
		$\Omega$  & maximum number of hours each trip can be delayed  \\
		$s_t$, $e_t$  & start and end hour of trip $t$ \\[8 pt]
	\end{tabularx}
\end{minipage}
\\
\begin{minipage}{\columnwidth}
	\noindent \textit{Variables} \\
	\setlength{\parindent}{0em}
	\indent
	\begin{tabularx}{\textwidth}{p{2.2 cm}X}
		$T_{it}$		&  1 if trip $t$ is driven by engineer $i$, 0 otherwise. \\
		$X_{ih}$	&  1 if engineer $i$ has a call window starting in hour $h$, 0 otherwise. \\
		$Y_{ih}$	&  1 if engineer $i$ has a call window covering hour $h$, 0 otherwise. \\
		$Z_{id}$	&  1 if engineer $i$ has a call window starting on day $d$, 0 otherwise. \\
	\end{tabularx}
\end{minipage}
\\
\\
\\
\noindent \textit{Call Window Construction Model for the set of new engineers ($I_b$) added to base $b$} 

\begin{subequations} \label{constructcwmodel}
	\begin{alignat}{3}
	\text{Maximize} \qquad \sum\limits_{i \in I_b} \sum\limits_{t \in T^{\text{U}}_b}  T_{it} + \sum\limits_{i \in I_b} \sum\limits_{d \in D} Z_{id} \label{eq:constructcwmodel_obj}
	\end{alignat}
	Subject to 
	\begin{enumerate}[itemsep=-1mm]
		\item Each trip $t$ can be covered by at most one engineer.
		\begin{alignat}{3}
		\sum\limits_{i \in I_b} T_{it} \leq 1  \qquad \qquad \forall \; t \in T^{\text{U}}_b  \label{eq:constructcwmodel_T1} 
		\end{alignat}
		
		\item \indent Engineer $i$ has a call window starting on day $d$ if and only if she has a call window starting in hour $h \in H_d$.
		\begin{alignat}{3}
		Z_{id} = \sum\limits_{h \in H_d} X_{ih}  \qquad \qquad \forall \; i \in I_b, \; d \in D  \label{eq:constructcwmodel_ZXdayhour} 
		\end{alignat}
		
		\item \indent If engineer $i$ has a call window starting in hour $h$, then $Z_{i,d_h}=1$.
		\begin{alignat}{3}
		Z_{i,d_h} \geq X_{ih}   \qquad \qquad \forall \;  i \in I_b, \; h \in H \label{eq:constructcwmodel_ZXhourday} 
		\end{alignat}
		
		\item \indent If engineer $i$ has a call window starting in hour $h$, then she will be on call during the period from hour $h$ to hour ($h+l^{\text{window}}-1$), and stop being on call in hour ($h+l^{\text{window}}$).
		\begin{alignat}{3}
		(l^{\text{window}} - \sum\limits_{h'=h}^{h+l^{\text{window}}-1} Y_{ih'}) + Y_{i,h+l^{\text{window}}}  \leq  (l^{\text{window}}+1) (1 - X_{ih})  \quad \forall \;i \in I_b, \;  h \in H \backslash LH_{l^{\text{window}}}    \label{eq:constructcwmodel_cwlength} 
		\end{alignat}
		
		\item \indent If hour $h$ is covered by a call window, then engineer $i$ must start a call window in exactly one hour between hours $h-l^{\text{window}}+1$ and $h$. Similarly, if hour $h$ is not covered by any call window, then engineer $i$ cannot be assigned a call window that starts between hours $h-l^{\text{window}}+1$ and $h$.
		\begin{alignat}{3}
		\sum\limits_{h'=h-l^{\text{window}}+1}^{h} X_{ih'} = Y_{ih}    \qquad \qquad \forall  \; i \in I_b, \; h \in H \backslash FH_{{l^{\text{window}}} } \label{eq:constructcwmodel_XY} 
		\end{alignat}
		
		\item \indent An engineer cannot be on call for more than $c^{\text{day}}$ days in a row.
		\begin{alignat}{3}
		\sum\limits_{d'=d}^{d+c^{\text{day}}} Z_{id'} \leq c^{\text{day}}   \qquad \qquad \qquad \forall \; i \in I_b, \; d \in D \backslash LD_{c^{\text{day}}}    \label{eq:constructcwmodel_consecutiveday}
		\end{alignat}
		
		\item \indent An engineer must have at least $\underline{d}^{\text{rest}}$ not-on-call days in the planning horizon.
		\begin{alignat}{3}
		\sum\limits_{d \in D} (1 - Z_{id}) \geq \underline{d}^{\text{rest}} \qquad \qquad \forall \; i \in I_b  \label{eq:constructcwmodel_nocwday}
		\end{alignat}
		
		\item \indent Engineer $i$ must have the same call window schedules on each on call day. If engineer $i$ has call windows on days $d$ and $d+n$, then the call window times on these two days must be exactly the same.
		\begin{alignat}{3}
		|X_{i,h+24n} -  X_{i,h}| \leq 2 - (Z_{id} + Z_{i,d+n}) \qquad \qquad  \forall \; i \in I_b, \;  d \in D \backslash LD_n, \; h \in H_d \backslash LH_{24n}, n \geq 1  \label{eq:constructcwmodel_samecw} 
		\end{alignat}
		
		(To linearize, constraints \eqref{eq:constructcwmodel_samecw} we replace them with the following.
		\begin{alignat}{3}
		- 2 + (Z_{id} + Z_{i,d+n}) \leq X_{i,h+24n} - X_{i,h} \leq 2 - (Z_{id} + Z_{i,d+n}) \qquad \qquad \qquad \qquad \qquad \qquad \nonumber \\
		\qquad \qquad \qquad \qquad \qquad \forall \; i \in I_b, \;  d \in D \backslash LD_n, \; h \in H_d \backslash LH_{24n}, n \geq 1)  \label{eq:constructcwmodel_samecw_linear} 
		\end{alignat}
		
		\item \indent Each engineer $i$ can have at most one call window starting on each day $d$.
		\begin{alignat}{3}
		\sum\limits_{h \in H_d} X_{ih} \leq 1 \qquad \qquad \forall \; i \in I_b, \; d \in D  \label{eq:constructcwmodel_onecwday}
		\end{alignat}
		
		\item \indent Engineer $i$ cannot be assigned trip $t$ (which is delayed by $\omega$ hours) if he is not on call in hour ($s_t + \omega - \delta_{\text{H}}$).
			\begin{alignat}{3}
			\sum\limits_{t \in   \bigcap\limits_{\omega=0}^\Omega  {  \{t|h \leq s_t + \omega < \delta_{\text{H}}+h, t \in T^{\text{U}}_b\}} } T_{it} \leq (\Omega+1) |T^{\text{U}}_b| \cdot Y_{ih}  \qquad \qquad \qquad \qquad \forall \; i \in I_b, \; h \in H \label{eq:constructcwmodel_advance_home}
			\end{alignat}

		\item Variables definitions
		\begin{alignat}{3}
			T_{it}, X_{ih}, Y_{ih}, Z_{id} \in \{0,1\} & \qquad \qquad \qquad \quad \forall \; i \in N, \; d \in D, \; h \in H, \; t \in T^{\text{U}}_b \label{eq:constructcwmodel_vardef}
		\end{alignat}
	\end{enumerate}
\end{subequations}

\indent Objective function \eqref{eq:constructcwmodel_obj} consists of two terms. The first  is intended  to maximize the number of covered trips; the second  aims to assign as many call windows as possible. Constraints \eqref{eq:constructcwmodel_ZXdayhour} require that if and only if engineer $i$ has a call window starting in an hour on day $d$, then logically, he has a call window starting on day $d$. Constraints \eqref{eq:constructcwmodel_ZXhourday} force $Z_{i,d_h}$ to be 1 if engineer $i$ has a call window starting in hour $h$, where $d_h$ is the day to which hour $h$ belongs. Constraints \eqref{eq:constructcwmodel_cwlength} restrict the call window length to be exactly $l^{\text{window}}$ hours. 

Constraints \eqref{eq:constructcwmodel_XY} require that if hour $h$ is covered by a call window, then engineer $i$ must start a call window in exactly one of the  hours between $h-l^{\text{window}}+1$ and $h$. These  constraints define the relationship between the $X$ and $Y$ variables: if hour $h$ is not in engineer $i$'s call window, then he must not start any call windows between hours $h-l^{\text{window}}+1$ and $h$. As such, if the engineer does not have a call window starting at a particular hour, then he will not be on call in the following $l^{\text{window}}-1$ hours. Constraints \eqref{eq:constructcwmodel_consecutiveday} add an upper bound on the consecutive number of on call days for engineer $i$. 

Constraints \eqref{eq:constructcwmodel_samecw} require that engineer $i$ must be assigned the same call window on each on call day. Constraints \eqref{eq:constructcwmodel_onecwday} require that engineer $i$ should have no more than one call window starting on each day. In Constraints \eqref{eq:constructcwmodel_advance_home}, if trip $t$ is assigned to engineer $i$, then he must be on call at least  $\delta_{\text{H}}$ hours before the trip departs. Constraints \eqref{eq:constructcwmodel_vardef} define the variables.

\section{Computational Experiments for Methodology 1: Set-covering-type Model} 
\label{sec_compexp_method1}

\noindent This section presents the computational experiments for the Set-covering-type Model. Section \ref{sec_compexp_design} introduces the experimental design and the simulation logic for making trip-engineer assignments. Section \ref{sec_compexp_dataparam} clarifies the data and parameter values to be used in the experiments. In Section \ref{sec_compexp_result}, we discuss the results and analyze the impact of different parameters on each cost metric for both 2-city and 3-city instances. In Section \ref{sec_compexp_sensitivity}, we conduct sensitivity analysis on the upper bound for the number of engineers.

\subsection{Design of Simulation of Call Window Performance} \label{sec_compexp_design}

\noindent In the formulation of Model \eqref{windowmodel}, it was assumed that all trip start times and travel times are deterministic. Under these conditions, solutions can be obtained in a matter of minutes with a commercial optimizer. In reality, though, a variety of disruptions occur daily, so the number of engineers and their call windows derived by solving  Model \eqref{windowmodel} may greatly underperform expectations.  To account for this uncertainty, we treat inter-departure times and trip lengths as random variables and sample their values from distributions derived from historical data (i.e., perform Monte Carlo simulation). For each random instance, we use discrete-event simulation (DES) to model the operational environment (i.e., the Board) in which trips are assigned to drivers. In this simulation, trip-engineer assignments are made on a first-in, first-out basis using the derived call windows.

Performance is evaluated by several key metrics including number of engineers required, the undercoverage rate, the percentage of engineers with 48 hours of consecutive rest, and the number of trips delayed. The  framework for the computations is as follows. \\[-18pt]

\begin{enumerate}
	\item Create an instance by randomly generating a 14-day train schedule in which inter-departure times and trip lengths are sampled from  given distributions. \label{step_computation_design_data} \\[-20pt]
	\item  Solve the Set-covering-type Model \eqref{windowmodel} given in Section \ref{sec_method} to determine the call window schedules in the current instance for the initial set of engineers. \label{step_computation_design_cw} \\[-20pt]
	\item Run the discrete-event simulation of the Board to assign uncovered trips to the engineers whose call windows were just obtained. The DES is highlighted in Figure \ref{Fig_FlowChart_Alg3} and Algorithm \ref{alg_simulation_itr} below.\label{step_computation_design_assign} \\[-20pt]
	\item If all trips are covered, or the number of engineers has reached its upper limit, then stop and go to Step \ref{step_computation_design_cost}. Otherwise, augment the set of engineers and solve Model  \eqref{windowmodel} using the set of uncovered trips only as input to get call windows for the new engineers. Go to Step \ref{step_computation_design_assign}.  \label{step_computation_design_stop} \\[-20pt]
	\item Calculate and report the resulting metrics. \label{step_computation_design_cost} 
\end{enumerate}

\indent The next subsections describe how the simulation assigns trips to engineers on the Board (Step \ref{step_computation_design_assign}). Sections \ref{sec_compexp_simu_2city} and \ref{sec_compexp_simu_3city} introduce the mechanism for 2-city and 3-city problems, respectively.

\subsubsection{Trip-engineer assignment simulation for 2-city problems} \label{sec_compexp_simu_2city}

\noindent This subsection introduces Algorithm \ref{alg_simulation_itr} that outlines trip-engineer operations for 2-city problems. In each iteration, Model \eqref{windowmodel} is solved to get call window schedules for the current set of engineers. Then, we assign the next trip to either (i)  the engineer who has been waiting longest on the Board at his home base, or (ii)  the engineer who has been waiting longest on the Board at his away base. Note that all HOS regulations in Section \ref{sec_probdes_assum} must be satisfied when a trip is assigned to an engineer. 

\indent The specific logic of the algorithm and the trip-engineer assignments is as follows. Among all the uncovered trips, we first find trip $t^*$ with the earliest start time. Then we delay trip $t^*$ by $\omega$ hours. To minimize delay, we start with $\omega=0$ and increase its value up to 2 until we find an eligible engineer to drive $t^*$. If no candidate is found, we search in the pool of engineers $I$ to check if there are any who can drive $t^*$ without violating the HOS constraints. Note that engineers in the set $I$ can have their home base in either city. 
	
\indent Let $I^{\text{home}}$ and $I^{\text{away}}$ be the sets of engineers who can drive $t^*$ without violating HOS constraints such that the origin of trip $t^*$ is their home base and away base, respectively. There are three possibilities:

\begin{enumerate}
	\item If $I^{\text{home}} \cup I^{\text{away}} = \phi$, it means there are no eligible engineers for $t^*$, so we stop. \\[-20pt]
	\item Else if $I^{\text{away}} \neq \phi $ (there are engineers whose home base is not the origin of $t^*$), then let the engineer in $I^{\text{away}}$ who has been waiting longest at the origin drive $t^*$ back to her home base. Let $i^*$ be the selected engineer. \\[-20pt]
	\item Else if $I^{\text{home}} \neq \phi$ (there are engineers whose home base is the origin of $t^*$), then let the engineer in $I^{\text{home}}$ who has been waiting longest at the origin drive $t^*$ to her away base. Let $i^*$ be the selected engineer. Next, we decide how  $i^*$ will return to her home: \\[-20pt]
	\begin{enumerate}
		\item If (i) the number of upcoming uncovered trips departing from $i^*$'s away base is greater than the number of engineers waiting longer than she and having home base where $t^*$ originates, and (ii) $i^*$ can rest at the destination of $t^*$ for 11.5-15 hours, then we let $i^*$ drive a trip back home if one is available. \\[-20pt]
		\item Otherwise, $i^*$ takes a van home (where van service is contracted to a local company and assumed to be always available).
	\end{enumerate}
\end{enumerate}

\begin{figure}[!h]
	\centering
	\includegraphics[width=7.3in]{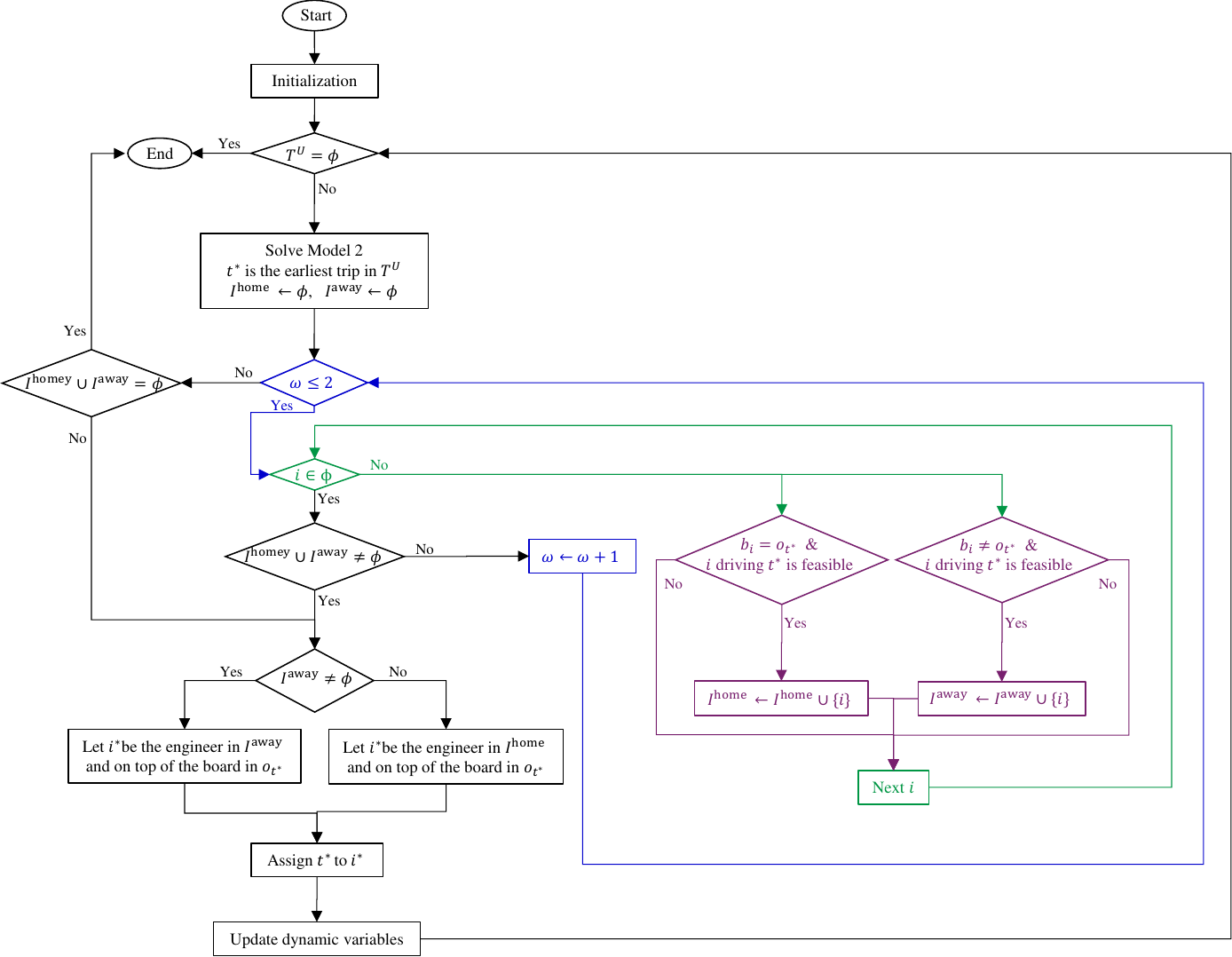}
	\caption{Computational flow of Algorithm \ref{alg_simulation_itr}}
	\label{Fig_FlowChart_Alg3}
\end{figure}

\indent If there are uncovered trips remaining, we go to the next iteration to generate additional call window schedules (for the new set of engineers introduced) by solving Model \eqref{windowmodel} with updated uncovered trips. The algorithm stops when all trips are covered or the maximum number of engineers has been scheduled. The following notation is used in the DES. In Algorithm \ref{alg_simulation_itr}, the complexities of initialization and iterative steps are $O(|I| \cdot |T^{\text{U}}|)$ and $O(|I| \cdot (|D|+|F|+|H|^2+|T|))$, respectively.  Figure \ref{Fig_FlowChart_Alg3} depicts the general computational flow. 
\\
\\
\begin{minipage}{\columnwidth}
	\noindent \textit{Sets and indices} \\
	\setlength{\parindent}{0em}
	\indent
	\begin{tabularx}{\textwidth}{p{2.8 cm}X}
		$b$		& index for base \\
		$B$		& set of bases that can potentially serve as the home base for engineers \\
		$b_1$, $b_2$	& two bases that can potentially serve as the home base; $B = \{b_1, b_2\}$ \\
		$d_{s_t}$, $d_{e_t}$ & index for the start and end day of train trip $t$ \\
		$d_{s_v}$, $d_{e_v}$ & index for the start and end day of van trip $v$ \\
		$f$		& index for period. For a planning horizon of 14 days as used in this study,  there are two periods: the first and the second 7 days. \\
	\end{tabularx}
\end{minipage}
\begin{minipage}{\columnwidth}
	\setlength{\parindent}{0em}
	\indent
	\begin{tabularx}{\textwidth}{p{2.8 cm}X}
		$F$		& set of periods \\
		$f_t$   & the period when trip $t$ starts \\
		$h_0$   & index for the first hour in the planning horizon \\
		$I$		& set of engineers. Each derived call window schedule is assigned to exactly one engineer. Two or more engineers can have the same call window schedule. \\
		$o_v$, $d_v$	& origin (destination) of van trip $v$; $o_v, \; d_v \in \{b_1, \; b_2\}$ \\
		$v$		& index for trip by van \\
		$V$		& set of trips by van \\
	\end{tabularx}
\end{minipage}
\\
\\
\\
\begin{minipage}{\columnwidth}
	\noindent \textit{Parameters from the data set} \\
	\setlength{\parindent}{0em}
	\indent
	\begin{tabularx}{\textwidth}{p{2.8 cm}X}
		$D_1$ ($D_2$) 	& maximum number of days an engineer can work in a row before given $r_1$  ($r_2$) hours consecutive rest at home base\\
		$k^{\text{DT}}_{dt}$ ($k^{\text{HT}}_{ht}$)		&  1 if trip $t$ covers day $d$ (hour $h$), 0 otherwise\\[7 pt]
		$k^{\text{DT}}_{dg}$ ($k^{\text{HT}}_{hg}$	)	&  1 if deadhead trip $g$ by van covers day $d$ (hour $h$), 0 otherwise \\[7 pt]
		$k^{\text{DH}}_{dh}$	&  1 if day $d$ covers hour $h$, 0 otherwise \\
		$l_t$, $l_v$  &  length of trip $t$ or $v$, $l_v = e_v - s_v$, $l_t = e_t - s_t$ \\
		$r_1$ ($r_2$)	&  minimum rest time after an engineer has worked $D_1$ ($D_2$) consecutive days \\
	\end{tabularx}
\end{minipage}
\begin{minipage}{\columnwidth}
	\begin{tabularx}{\textwidth}{p{2.8 cm}X}
		$r^{\text{trip}}$	&  minimum rest time between driven trips \\
		$r^{\text{period}}$	&  minimum number of consecutive rest hours we try to provide for each engineer in a period \\
		$s_v$, $e_v$     & start and end time of van trip $v$  \\
		$w^{\text{period}}$	&  maximum number of hours an engineer can drive in a period \\
		\end{tabularx}
\end{minipage}
\\
\\
\begin{minipage}{\columnwidth}
	\noindent \textit{Parameters from call window schedules obtained by Model \eqref{windowmodel}} \\
	\setlength{\parindent}{0em}
	\indent
	\begin{tabularx}{\textwidth}{p{2.8 cm}X}
		$cd_{id}$ & number of consecutive on call days of engineer $i$ up until day $d$; if the engineer is not on call on day $d$, then $cd_{id}=0$ \\
		$oc_i (p_1, p_2)$  & (binary)  1 if engineer $i$ is on call between time points $p_1$ and $p_2$, 0 otherwise \\
		$wh_{if}$  & number of on call hours for engineer $i$ in period $f$ \\
	\end{tabularx}
\end{minipage}
\\
\\
\\
\begin{minipage}{\columnwidth}
	\noindent \textit{Dynamic variables and sets} \\
	\setlength{\parindent}{0em}
	\indent
	\begin{tabularx}{\textwidth}{p{2.8 cm}X}
		$p$				&  time point in the planning horizon; i.e., the number of hours and fractions thereof from the beginning of the planning horizon (For example, assume that each day starts at 7:00 am. If the time point is 8:30 am on day 2, then $p=25.5$.)   \\
		$T^{\text{U}}$    & set of uncovered trips \\
		$X_{it}$   & (binary variable)  1 if engineer $i$ drives trip $t$, 0 otherwise \\
		$CD_{id}$ & number of consecutive working days of engineer $i$ up until day $d$; if the engineer does not work on day $d$, then $CD_{id}=0$ \\
		$L_{ip}$ & location of engineer $i$ at time point $p$; $L_{ip}$ can be $b_1$, $b_2$ or trip $t$ \\
		$WH_{if}$  & number of working hours of engineer $i$ in period $f$ \\
		$CR^1_i(p_1, p_2)$ & (binary variables) equal 1 if engineer $i$ is off duty between time points $p_1$ and $p_2$  during a mandatory $r_1$-hour rest period, 0 otherwise (in the computations, $r_1 = 48$) \\
		$CR^2_i(p_1, p_2)$  &  (binary variables) equal 1 if engineer $i$ is off duty between time points $p_1$ and $p_2$ during a mandatory  $r_2$-hour rest period, 0 otherwise (in the computations, $r_2 = 72$)\\
		$OC_i (p_1, p_2)$  & (binary variable)  1 if engineer $i$ is on call between time points $p_1$ and $p_2$, 0 otherwise \\
	\end{tabularx}
\end{minipage}

\begin{algorithm}
	\begin{algorithmic}[1]
		\caption{DES (initialization and iterative steps) of the current trip-engineer assignment system with call window schedules from Model \eqref{windowmodel}} \label{alg_simulation_itr}
		\State \textbf{\textit{Initialization}}
		\State $T^{\text{U}} \leftarrow T$, $T^{\text{E}} \leftarrow \phi$
		\For{$i \in I$}
		\For{$p, p' \in H$}
		$CR^1_i (p, p') \leftarrow 0$, \; $CR^2_i (p, p')\leftarrow 0$, \; $OC_i (p, p') \leftarrow oc_i (p, p')$
		\EndFor
		\For{$d \in D$}
		$CD_{id} \leftarrow cd_{id}$
		\EndFor
		\For{$f \in F$} 
		$WH_{if} \leftarrow wh_{if}$
		\EndFor
		\For{$t \in T$}  
		$X_{it} \leftarrow 0$
		\EndFor
		\EndFor	
		
		\State \textbf{\textit{Iterative Steps}}
		\While{$T^{\text{U}} \neq \phi$}
		\State Solve Model \eqref{windowmodel} with demand from $T^{\text{U}}$ to get call window schedules
		\State $t^* \leftarrow \text{argmin}\{t|s_t, t \in T^{\text{U}}\} $, \; $p \leftarrow s_{t^*}$, \; $I^{\text{home}} \leftarrow \{\}$, \; $I^{\text{away}} \leftarrow \{\}$
		\For{$\omega \in \{0,1,2\}$} $s_{t^*} \leftarrow s_{t^*} + \omega$
		\For{$i \in I$}
		\If{(1) $ \exists \; (p_1, p_2) \text{ s.t. } p_1 < p < p_2, \; OC_i (p_1, p_2) = 1$ ($i$ is on call)  \\
			\qquad \qquad \qquad \&  (2) $\nexists \; (p_1, p_2) \text{ s.t. } p_1 < p < p_2, \; CR^1_i(p_1, p_2) + CR^2_i(p_1, p_2) \geq 1$ \\
			\qquad \qquad \qquad \qquad \quad ($i$ is not in $r_1$ or $r_2$-hour rest) \\
			\qquad \qquad \qquad \&  (3) $CD_{i d_{s_t}} + (d_{e_v} - d_{s_t}) < D_2$ ($i$ will not work more than $D_2$ days in a row) \\
			\qquad \qquad \qquad \&  (4) $WH_{if_{t*}} + l_t + l_v < w^{\text{period}}$ ($i$ works no more than $ w^{\text{period}}$ hours in period $f_t$) \\
			\qquad \qquad \qquad \&  (5) \& $L_{i, s_{t^*}} = o_{t^*}$ ($i$ is in the origin of $t^*$ when $t^*$ departs)} 
		\If{$b_i = o_{t^*}$ ($t^*$'s origin is $i$'s home base) \\
			\qquad \qquad \qquad \qquad \& $ \exists \; (p_1, p_2) \text{ s.t. } p_1 < p - \Delta^{\text{h}} < p_2, \; OC_i (p_1, p_2) = 1$ \\
			\qquad \qquad \qquad \qquad \quad ($i$ was on call $\Delta^{\text{h}}$ hours ago)}  
		\qquad \qquad \qquad \qquad \qquad \qquad \State $I^{\text{home}} \leftarrow I^{\text{home}} \cup \{i\}$
		\ElsIf{$b_i = d_{t^*}$ ($t^*$'s origin is $i$'s away base) \\
			\qquad \qquad \qquad \qquad \qquad \& $ \exists \; (p_1, p_2) \text{ s.t. } p_1 < p - \Delta^{\text{a}} < p_2, \; OC_i (p_1, p_2) = 1$ \\
			\qquad \qquad \qquad \qquad \qquad \quad  ($i$ was on call $\Delta^{\text{a}}$ hours ago)}
		\State $I^{\text{away}} \leftarrow I^{\text{away}} \cup \{i\}$
		\EndIf
		\EndIf
		\EndFor
		\If{$I^{\text{away}} \cup I^{\text{home}} \neq \phi$} break
		\EndIf
		\EndFor
		\If{$I^{\text{away}} \neq \phi$}
		\State Let $i^*$ be the engineer who is in $I^{\text{away}}$ and on top of the board in base $b$ 
		\State $X_{i^* t^*} \leftarrow 1$ (assign $t^*$ to $i^*$)
		\ElsIf{$I^{\text{home}} \neq \phi$}
		\State Let $i^*$ be the engineer who is in $I^{\text{home}}$ and on top of the board in base $b$ 
		\State $X_{i^* t^*} \leftarrow 1$ (assign $t^*$ to $i^*$)
		\If{(the number of trips from $b'$ is 11.5-15 hours after $e_t$) \\
			\qquad \qquad  $\geq$ 1 + (the number of ``away" engineers in $b'$)}
		\State Let $i^*$ rest in $b'$ and drive another train back home
		\Else let $i^*$ take van home
		\EndIf
		\EndIf
		\State Update variables $CD$, $WH$, $T^{\text{U}}$,  $OC$, $L$		
		\EndWhile

		\State \textbf{\textit{Calculate Cost Metrics}}
		\State Calculate: number of trips by eBoard engineers, number of working engineers, number of engineers with at least 48-hours of consecutive rest in one week, average number of trips and working/driving hours for each engineer, percentage of trips driven from engineers' home base, number of delayed trips per week, and number of delayed hours for each trip.
	\end{algorithmic}
\end{algorithm}

\subsubsection{Trip-engineer assignment simulation for 3-city problems}  \label{sec_compexp_simu_3city}

\noindent In this section, we introduce the algorithm to construct call window schedules and manage the Board for 3-city problems. Assume we have three cities ($b_1$, $b_2$,  $b_3$) in the district. There are trips between $b_1$ and $b_2$, $b_2$ and $b_3$, but not between $b_1$ and $b_3$. Engineers with home base in $b_2$ can drive any trips in the system. If an engineer's home is in $b_1$, then he can only drive between $b_1$ and $b_2$. Similarly, if an engineer's home is in $b_3$, then he can only drive between $b_2$ and $b_3$.

\indent The algorithm to construct call window schedules and make trip-engineer assignments for 3-city problems is similar to that for 2-city problems. Specifically, for each city $b$, we use Model \eqref{windowmodel} to assign call window schedules to the available engineers residing in base $b$. The next step is to run Algorithm \ref{alg_simulation_itr} to assign trips to the engineers. In Algorithm \ref{alg_simulation_itr}, the difference between 2-city and 3-city problems lies in line 24. For 2-city problems, the logic in line 24 considers only one away base for engineer $i$, while for 3-city problems, line 24 considers two away bases for engineer $i$.  As such, when $I^{\text{away}} \neq \phi$, the engineers in $I^{\text{away}}$ originate from two home bases, so among the engineers with different home bases, we select the one who has been waiting longest at the origin of trip $t^*$ and let her drive $t^*$ back home.

\subsection{Data set and parameters} 
\label{sec_compexp_dataparam}

\noindent Based on our analysis of trip data provided by a Class I freight rail company, we were able to fit the  inter-departure times to an exponential distribution but had to resort to an  empirical distribution for trip lengths. The data sets used in the study were randomly generated by Algorithm \ref{alg_simulation_datagenerate} using these distributions.  Its complexity is $O(|B|\cdot |H|)$. 

\begin{algorithm}
	\begin{algorithmic}[1]
		\caption{Trip data generation} \label{alg_simulation_datagenerate}
		\State \textbf{\textit{Data Generation}} 
		\For{$b \in B$}
		$p = 0$
		\While{$p < |H|$}
		\State Generate trip $t$ that satisfies the following conditions: 
		\State \qquad (1) The origin of $t$ is $b$ 
		\State \qquad (2) The inter-departure time is generated from exponential distribution. 
		\State \qquad \quad \; The trip start time is $s$.
		\State \qquad (3) The trip length is generated from an empirical distribution.
		\State \qquad (4) If the generated trip length is greater than 12, then we fix its length to 12 hours
		\State $p \leftarrow p + s$
		\EndWhile
		\EndFor
	\end{algorithmic}
\end{algorithm}

\indent Each trip is defined by its origin-destination, trip start time and length. More specifically, the trip length is generated from an empirical distribution with a mean of 8.37 hours;  inter-departure times are generated from an exponential distribution with a mean of 8.97 hours. For the 2-city instances, the number of trips over a 4-week planning horizon ranged from 139 to 192 with an average of 168. For the 3-city instances, the number of trips was between 325 and 440 with an average of 380.

\indent Parameters for operations and HOS regulations (such as minimum rest time between two trips and maximum driving time in a week) are the same as those in Sections \ref{sec_probdes_cons} and \ref{sec_probdes_assum}.  Each return trip by van was assumed to be 3 hours and the penalty weight for each uncovered trip was set to 5. The latter value  turned out to be the best compromise after running dozens of trials.

\indent The upper bound on the number of engineers was calculated by multiplying the upper bound derived from the methodology used in the study of Guo and Bard (2024) by a parameter $\beta$. The introduction of $\beta$ provides a buffer to account for disruptions.   When the number of engineers reaches the upper bound, the algorithm stops. In the experiments, we start with  $\beta =1.2$ and then explore smaller values.

\indent To be able to draw statistical conclusions, we generated 25 instances for both 2-city and 3-city problems. In each instance, we ran experiments with the 24 different combinations of the following parameters.
\begin{enumerate}
	\item Call window length: 6, 10, 12, 24 hours \\[-20pt]
	\item Maximum number of delayed hours for each trip: 0, 1, 2  \\[-20pt]
	\item Always require engineers to return to their home base by van: yes, no
\end{enumerate}

\indent All algorithms were coded in Python and run on a Lenovo laptop with a 64-bit operating system, an Intel(R) Core(TM) i5-7200U CPU 2.50 GHz and 8.00 GB of RAM. CPLEX 22.1.1 was called to solve all mixed-integer linear programs. The remainder of this section discusses the values of the cost metrics obtained by solving for each combination of parameters listed in Section \ref{sec_compexp_dataparam}.

\subsection{Computational Results} 
\label{sec_compexp_result}

\noindent This subsection investigates the impact of call window length, the maximum number of delayed hours, and whether to allow engineers to drive a train home on each cost metric. Both 2-city and 3-city instances are discussed with $\beta=1.2$. 

\indent  In  Algorithm \ref{alg_simulation_itr}, trips are assigned to engineers in a myopic (greedy) manner. While this strategy does not guarantee  globally optimal assignments, it provides a computationally efficient method to generate feasible schedules. Naturally, engineers added in later iterations cover fewer trips than those assigned earlier, as the most favorable trips are assigned first. This outcome reflects the trade-off between prioritizing tractability and speed over global optimality. Nevertheless, the approach is arguably justified in our context, as it enables timely analysis while still capturing the main operational patterns. 

\subsubsection{2-city instances}

\noindent Table \ref{tbl_2city_summary} in Appendix \ref{appendix_results_2/3city} gives the average statistics obtained for the 25 instances with 2 cities.  In each instance, we ran 24 experiments with different combinations of parameters.  The remainder of this subsection discusses how each parameter impacts the cost metrics. In the statistical tests, if the p-value is smaller than 0.001, then we round down to 0 for simplicity. \\ [-10pt]

\noindent \textit{\textbf{Impact of the call window length.}} Table \ref{tbl_2city_length} in Appendix \ref{appendix_results_2/3city} presents the means and ANOVA p-values of costs for each call window length. It can be seen that a longer call window leads to the following outcomes.
\begin{enumerate}[label=(\roman*)]
	\item A lower undercoverage rate (p-value 0) \\[-20pt]
	\item Insignificantly fewer engineers (p-value  0.353) \\[-20pt]
	\item Fewer engineers with at least one 48-hour consecutive rest per week (p-value 0)  \\[-20pt]
	\item Larger workload per week (a greater number of trips, driving hours and working hours with p-values of 0) \\[-20pt]
	\item A higher percentage of trips driven by engineers from their home base (p-value 0) \\[-20pt]
	\item Insignificant change in delay (average number of delayed trips per week, number of delayed hours per delayed trip, and number of delayed hours for all trips, with p-values of 0.1654, 0.9992 and 0.3453, respectively)
\end{enumerate}

\noindent \textit{\textbf{Impact of the maximum number of delayed hours.}}  Table \ref{tbl_2city_delayhour} in Appendix \ref{appendix_results_2/3city} presents the means and ANOVA p-values for costs for each upper limit on the number of delayed hours. The results show that a higher upper limit on the number of delayed hours leads to following results.
\begin{enumerate}[label=(\roman*)]
	\item A lower undercoverage rate (p-value 0) \\[-20pt]
	\item Insignificantly fewer working engineers (p-value  0.9255) \\[-20pt]
	\item Insignificant impact on the number of engineers with at least one 48-hour consecutive rest per week (p-value 0.2767) \\[-20pt]
	\item Greater workload per week (a greater number of trips, driving hours and working hours with p-values of 0.0051, 0.0091 and 0.0076, respectively) \\[-20pt]
	\item Insignificant impact on the percentage of trips driven by engineers from their home base (p-value 0.9841) \\[-20pt]
	\item More serious delay (a larger number of delayed trips per week, and average number of delayed hours for each trip, with p-values of 0) \\[-20pt]
\end{enumerate}

\noindent \textit{\textbf{Impact of whether to force engineers to return home by van.}} Table \ref{tbl_2city_YN} in Appendix \ref{appendix_results_2/3city} presents the means and paired t-test p-values for costs whether or not the engineers are required to always return home by van (Y), or to drive a train home (N). According to the results, we see that allowing engineers to return home by driving a train creates:
\begin{enumerate}[label=(\roman*)]
	\item A lower undercoverage rate (p-value 0) \\[-20pt]
	\item Insignificant impact on the number of  engineers (p-value 0.6064) \\[-20pt]
	\item Fewer engineers with at least one 48-hour consecutive rest per week (p-value 0) \\[-20pt]
	\item Greater workload per week (a greater number of trips, driving hours and working hours with p-values of 0) \\[-20pt]
	\item Lower percentage of trips driven by engineers from their home base (p-value 0) \\[-20pt]
	\item Less serious delays (a smaller number of delayed trips per week, and average number of delayed hours for all trips, with p-values of 0), and insignificant impact on the average number of delayed hours for each delayed trip (p-value 0.4797) \\[-20pt]
\end{enumerate}

One observation from the results is that the number of trips assigned to each engineer is small, i.e., each engineer drives approximately 2.22 trips per week. The reason for the low workload is the additional duties that engineers expected to do centered on safety and operational efficiency. These involve meticulous pre-trip inspections of equipment and brakes, fielding dispatcher requests, and continuous monitoring track conditions and onboard systems. Furthermore, drivers are responsible for managing train handling to protect cargo, assisting in switching cars in the yard, repositioning equipment in the station, and serving as a first responder for any emergencies or mechanical issues that arise. 

\subsubsection{3-city instances}

\noindent The discussion of the computational results for 3-city instances is given in Appendix \ref{appendix_results_discussion_method1_3city}.

\subsection{Sensitivity Analysis on Buffer Value $\beta$} \label{sec_compexp_sensitivity}

\noindent In Section \ref{sec_compexp_result}, the computations reported were derived with $\beta=1.2$. In this section, we set $\beta=1.0$ and investigate the  impact of this reduction on the cost metrics.

\subsubsection{Impact of $\beta$ on 2-city Instances}

\noindent Table \ref{tbl_2city_buff} presents the average cost values obtained with the set-covering-type model. The second and third columns display the mean costs with $\beta=1.0$ and $1.2$, respectively. These values were obtained by calculating the average of all 600 individual cases, where $600=(25 \text{ instances}) \times (4 \text{ call window lengths}) \times (3 \text{ maximum number of delayed hours}) \times (2 \text{ options for returning home})$. The fourth column is the difference in means between the two $\beta$ values. The difference equals the cost with $\beta=1.2$ minus that with $\beta=1.0$. The last column is the p-value from the paired t-test. A p-value below 0.05 indicates the difference in the means between $\beta=1.0$ and $1.2$ is significant.

\renewcommand{\arraystretch}{1.3}
\begin{table}[htbp]
	\centering
	\caption{Means and difference in costs for 2-city instances with $\beta=1.0$ and $\beta=1.2$ obtained from the Set-covering-type Model.}
	\resizebox{\columnwidth}{!}{\begin{tabular}{|l|c|c|c|c|c|}
			\hline
			\multicolumn{1}{|p{4.125em}|}{Cost}  & \multicolumn{1}{p{4.125em}|}{Mean ($\beta=1.0$)} &
			\multicolumn{1}{p{4.125em}|}{Mean ($\beta=1.2$)} & \multicolumn{1}{p{4.125em}|}{Difference (1.2 - 1.0)} & p-value \\ \hline
			Undercoverage rate (\%) & 10.98 & 5.11 & -5.87 & 0 \\
			Average number of working engineers & 15.44 & 18.02 & 2.58 & 0 \\
			Average number of engineers with at least one 48-hr rest in a row per week & 14.62 & 16.40 & 1.78 & 0 \\
			Average number of train trips for each engineer per week & 2.43 & 2.22 & -0.21 & 0 \\
			Average number of driving hours for each engineer per week & 20.21 & 18.45 & -1.75 & 0 \\
			Average number of working hours for each engineer per week & 27.50 & 25.12 & -2.38 & 0 \\
			Percentage of trips driven from engineer's home base (\%) & 96.36 & 96.34 & -0.02 & $0.4678$ \\
			Average number of delayed trips per week & 2.60 & 2.66 & 0.07 & 0 \\
			Average number of delayed hours for each delayed trip & 0.84 & 0.84 & 0.00 & $0.2727$ \\
			Average number of delayed hours for all trips & 0.08 & 0.08 & 0.00 & $0.0014$  \\ \hline
	\end{tabular}}
	\label{tbl_2city_buff}%
\end{table}%

\renewcommand{\arraystretch}{1.1}

\indent From Table \ref{tbl_2city_buff}, we see that a larger upper bound on the number of engineers produces the following results.

\begin{enumerate}
	\item A significantly lower undercoverage rate. This is not surprising because we have more manpower to cover trips when other parameters are unchanged.
	\item Significantly more working engineers. This is because a higher upper limit on  manpower allows the algorithm to select more engineers to cover trips.
	\item A significantly greater number of engineers with at least one consecutive 48-hour rest period per week. Since a greater number of engineers leads to a lower probability of manpower scarcity, engineers have a higher chance of  getting 48 hours of rest in a row.
	\item A significantly lower weekly workload. Based on the assumption that the number of trips is fixed, the more engineers we have, the fewer the number of trips each engineer needs to drive on average. Consequently, the number of driving and working hours decreases as well.
	\item An insignificant change in the percentage of trips driven from the engineers' home base. 
	\item A significantly larger number of delayed trips per week. This result is surprising. The number of delayed trips per week was expected to decrease as the upper bound on the number of engineers grows, but the result demonstrates the opposite. This is due to the  logic used in the operations center. In each iteration of the algorithm, we first solve the set-covering-type model to generate call window schedules for the added engineers; then we assign trips to on call engineers. By the end of the first iteration, if there are uncovered trips, we go to the next iteration. The algorithm stops when either all the trips are covered or no more engineers can be added (i.e., the number of engineers has reached its upper bound). A higher $\beta$ value can lead to more iterations. Even though a higher coverage rate can be achieved in additional iterations, extra delays may also be created.
	
	\hspace{0.5 in}  For example, assume the upper bounds on the number of engineers are 10 and 12 for $\beta=1.0$ and $1.2$, respectively. In the first iteration, the set-covering-type model for both $\beta=1.0$ and $1.2$  generates 10 engineers (Engineers 1, 2,...,10). The schedules for the two cases are exactly the same, indicating that they have the same number of delayed trips and delayed hours (assume 10 trips are delayed, and each is delayed by 1 hour). By the end of the first iteration, there are still some uncovered trips. 
	
	\hspace{0.5 in} For the case where $\beta=1.0$ , we cannot go to the second iteration because the number of existing engineers has reached its upper bound. For the $\beta=1.2$ case, we go to the second iteration and can add at most 2 engineers. The additional 2 engineers can not only cover some of the remaining trips, but they also create more delayed trips. For instance, Engineer 11 can cover trip $t$, but in the example,  trip $t$ is delayed by 1 hour to meet the engineer's availability. Therefore, when $\beta=1.2$, there are 11 delayed trips and  a total delay of 11 hours. When $\beta=1.0$, there are 10 delayed trips and a total delay of 10 hours. Thus,  the number of delayed trips is greater in the case of $\beta=1.2$.
	\item An insignificant change in the average number of delayed hours for each delayed trip. Let's still use the above example. When $\beta=1.0$, there are 10 delayed trips and the total delayed time is 10 hours, so the average number of delayed hours for each delayed trip is $\frac{10}{10}=1$. Similarly, when $\beta=1.2$, there are 11 delayed trips and the total delayed time is 11 hours, so the average number of delayed hours for each delayed trip is $\frac{11}{11}=1$. Thus, the difference in the average number of delayed hours for each delayed trip is insignificant.
	\item A significantly larger number of delayed hours for all trips. Still using the same example, assume there are a total of 100 trips.  When $\beta=1.0$, the total delayed time is 10 hours, so the average number of delayed hours for all trips is $\frac{10}{100}=0.1$. When $\beta=1.2$, the total delayed time is 11 hours, so the average number of delayed hours for all trips is $\frac{11}{100}=0.11$. Therefore, when $\beta=1.2$, the  average number of delayed hours for all trips is greater.
	
\end{enumerate}

\indent In addition to the change in cost metrics, the change in the impact of call window lengths, the maximum number of delayed hours, and whether to allow engineers to drive a train home is insignificant. Essentially, the impacts of those parameters on the cost metrics are same for $\beta=1.0$ and $\beta=1.2$.

\subsubsection{Impact of $\beta$ on 3-city Instances}
\noindent The sensitivity analysis of $\beta$ for 3-city instances is given in Appendix \ref{appendix_results_discussion_method1_3city}.

\section{Computational Experiments for Methodology 2: Direct Model} 
\label{sec_compexp_method2}

\noindent This section presents the computational results obtained from the Direct Algorithm. All data and parameters are the same as those in Section \ref{sec_compexp_dataparam}. Section \ref{sec_compexp_design_method2} introduces the experiment design. In Section \ref{sec_compexp_result_method2}, we discuss the results and analyze the impact of different parameter values on each cost metric for both 2-city and 3-city problems. In Section \ref{sec_compexp_sensitivity_method2}, we conduct a sensitivity analysis on the upper bound on the number of engineers.

\subsection{Design of Simulation of Call Window Performance from Method 2: Direct Algorithm} \label{sec_compexp_design_method2}

\noindent The details of assigning trips to engineers based on the call windows generated in Section \ref{sec_method2} are the same as those presented in Algorithm \ref{alg_simulation_itr} in Section \ref{sec_compexp_design}. Algorithm \ref{alg_DirectAlg} gives the pseudocode for generating call window schedules using our DES. The notation is the same as that in Section \ref{sec_compexp_design}. The algorithm for 2-city and 3-city problems are similar to the one in Sections \ref{sec_compexp_simu_2city} and \ref{sec_compexp_simu_3city}, i.e., after we select call window schedules, we assign the next trip to either the engineer who has been waiting longest on the Board (first in the queue) at his home base, or the engineer who has been waiting longest on the Board at his away base.

\begin{algorithm}
	\begin{algorithmic}[1]
		\caption{Algorithm (initialization and iterative steps) to construct call window schedules based on the current trip-engineer assignment process and Model \eqref{constructcwmodel}} \label{alg_DirectAlg}
		\State \textbf{\textit{Initialization}}
		\State $T^{\text{U}} \leftarrow T$, $T^{\text{E}} \leftarrow \phi$, $I \leftarrow \phi $
		
		\State \textbf{\textit{Iterative Steps}}
		\While{$T^{\text{U}} \neq \phi$}
		\For{$b \in B$}
		\State Let $T^{\text{U}}_b$ be the set of uncovered trips starting from base $b$
		\State Solve Model \eqref{constructcwmodel} with $T^{\text{U}}_b$ to get call window schedule $s$ for an engineer with home base $b$
		\State Assign call window schedule $s$ to engineer $i$
		\State $I \leftarrow \{i\} \cup I$ 
		\EndFor
		\State Apply the trip-assignment mechanism discussed in lines 11-39 of Algorithm \ref{alg_simulation_itr}	
		\EndWhile

		\State \textbf{\textit{Calculate Cost Metrics}}
		\State Calculate: number of trips by eBoard engineers, number of working engineers, number of engineers with 48-hour rest, average number of trips and working/driving hours for each engineer
	\end{algorithmic}
\end{algorithm}

\subsection{Computational Results for Direct Algorithm} 
\label{sec_compexp_result_method2}

\noindent This subsection investigates the impact of call window length, the maximum number of delayed hours, and whether to allow engineers to drive a train home on each cost metric. The focus is on the 2-city problems with $\beta =1.2$; results for the 3-city problems are discussed in Appendix \ref{appendix_results_discussion_method2_3city}.

\subsubsection{Results for 2-city Instances}

\noindent Table \ref{tbl_2city_summary_method2} in Appendix \ref{appendix_results_2/3city_method2} gives the average output statistics obtained from 25 instances with 2 cities.  In each instance, we ran 24 experiments with different combinations of parameters.  \\ [-10pt]

\noindent \textit{\textbf{Impact of the call window length.}} Table \ref{tbl_2city_length_method2} in Appendix \ref{appendix_results_2/3city_method2} presents the means and ANOVA p-values of costs for each call window length. Longer call windows have the following implications. 
\begin{enumerate}[label=(\roman*)]
	\item A lower undercoverage rate (p-value 0) \\[-20pt]
	\item Insignificantly fewer engineers (p-value 0.4930) \\[-20pt]
	\item Fewer engineers with at least one 48-hour consecutive rest per week (p-value 0) \\[-20pt]
	\item A greater workload per week (a greater number of trips, driving hours and working hours with p-values of 0) \\[-20pt]
	\item A lower percentage of trips driven by engineers from their home base (p-value 0) \\[-20pt]
	\item Less serious delays (smaller average number of delayed trips per week, and number of delayed hours for all trips, with p-values of 0), while the number of delayed hours for each delayed trip is not significantly impacted. \\[-20pt]
\end{enumerate}

\indent Note that for (i) and (iii)-(v), the cost metrics are monotone in call window length only when the call window length is between 6 and 12 hours (not extended to 24 hours). For example, the undercoverage rate is 12.51\%, 4.29\% and 2.49\% for 6-hr, 10-hr and 12-hr call window, respectively, but it increases to 4.88\% in the 24-hr instances. This is because in each iteration of Algorithm \ref{alg_DirectAlg}, for different call window lengths, Model \eqref{constructcwmodel} generates different call window schedules. Moreover, in the experiments, whether we can assign a trip to an engineer not only depends on his call window schedule, but also on the HOS constraints discussed in Sections \ref{sec_probdes_cons} and \ref{sec_probdes_assum}. Therefore, a schedule with  longer call windows cannot guarantee a lower undercoverage rate.   \\ [-10pt]

\noindent \textit{\textbf{Impact of the maximum number of delayed hours.}}  Table \ref{tbl_2city_delayhour_method2} in Appendix \ref{appendix_results_2/3city_method2} presents the means and ANOVA p-values for costs for each upper limit on the number of delayed hours. It can be seen that a higher upper limit on the number of delayed hours indicates the following.
\begin{enumerate}[label=(\roman*)]
	\item A lower undercoverage rate (p-value 0) \\[-20pt]
	\item Insignificantly fewer engineers (p-value 0.7299) \\[-20pt]
	\item A smaller number of engineers with at least one 48-hour consecutive rest period per week (p-value 0) \\[-20pt]
	\item A greater workload per week (a greater number of trips, driving hours and working hours with p-values of 0) \\[-20pt]
	\item An insignificant impact on the percentage of trips driven by engineers from their home base (p-value 0.9630) \\[-20pt]
	\item More serious delays (a greater number of delayed trips per week, and average number of delayed hours for each trip, with p-values of 0) \\[-20pt]
\end{enumerate}

\noindent \textit{\textbf{Impact of whether to require engineers return home by van.}} Table \ref{tbl_2city_YN_method2} in Appendix \ref{appendix_results_2/3city_method2} presents the means and paired t-test p-values of costs for the policy as to  whether engineers are required to return home by van (Y), or to allow them to drive a train home (N). The results indicate that allowing engineers to return home by driving a train leads to the following outcomes.
\begin{enumerate}[label=(\roman*)]
	\item An insignificant change in undercoverage rate (p-value 0.4586) \\[-20pt]
	\item Insignificant change in the number of  engineers (p-value 0.7395) \\[-20pt]
	\item Fewer engineers with at least one 48-hour consecutive rest period per week (p-value 0.0112) \\[-20pt]
	\item No significant impact on weekly workload (number of trips, driving hours and working hours with p-values of 0.5693, 0.4598 and 0.4823, respectively) \\[-20pt]
	\item A lower percentage of trips driven by engineers from their home base (p-value 0) \\[-20pt]
	\item A smaller average number of delayed hours for all trips (p-value 0.0403). The number of delayed trips per week, and the average number of delayed hours for each delayed trip are insignificantly impacted with p-values of 0.1725 and 0.3208, respectively. \\[-20pt]
\end{enumerate}

\subsubsection{Results of 3-city Instances}

\noindent The discussion of the computational results for 3-city instances is given in Appendix \ref{appendix_results_discussion_method2_3city}.

\subsection{Sensitivity Analysis on Buffer Value $\beta$ for Direct Algorithm} \label{sec_compexp_sensitivity_method2}

\noindent In this section, we investigate the impact of the $\beta$ value on the cost metrics. The results obtained with the Direct Algorithm for $\beta=1.0$ and  $\beta=1.2$ are compared.

\subsubsection{Impact of $\beta$ on 2-city Instances}

\noindent  The second and third columns of Table \ref{tbl_2city_buff_method2} display the results for $\beta=1.0$ and $1.2$, respectively. These values were obtained by calculating the average of all 600 individual cases, where $600=(25 \text{ instances}) \times (4 \text{ call window lengths}) \times (3 \text{ maximum number of delayed hours}) \times (2 \text{ options for returning home})$. The fourth column is the difference in means between the two values; i.e., the costs for $\beta=1.2$ minus those for $\beta=1.0$. The last column is the p-value from the paired t-test. A p-value below 0.05 indicates that the difference in the means is significant.

\begin{table}[htbp]
	\centering
	\caption{Means and difference in costs for 2-city instancess with $\beta=1.0$ and $\beta=1.2$ obtained from the Direct Algorithm.}
	\resizebox{\columnwidth}{!}{\begin{tabular}{|l|c|c|c|c|}
		\hline
		\multicolumn{1}{|p{4.125em}|}{Cost}  & \multicolumn{1}{p{4.125em}|}{Mean ($\beta=1.0$)} &
		\multicolumn{1}{p{4.125em}|}{Mean ($\beta=1.2$)} & \multicolumn{1}{p{4.125em}|}{Difference (1.2 - 1.0)} & p-value \\ \hline
		Undercoverage rate (\%) & 11.03 & 6.04  & -4.99 & 0 \\
		Average number of working engineers & 15.42 & 17.90 & 2.49 & 0  \\
		Average number of engineers with at least one 48-hr rest in a row per week & 14.44 & 15.99 & 1.54  & 0 \\
		Average number of train trips for each engineer per week & 2.43  & 2.21  & -0.22 & 0 \\
		Average number of driving hours for each engineer per week & 20.23 & 18.39 & -1.84 & 0 \\
		Average number of working hours for each engineer per week & 27.53 & 25.04 & -2.50 & 0 \\
		Percentage of trips driven from engineer's home base (\%) & 98.93 & 99.00 & 0.06  & 0.0010 \\
		Average number of delayed trips per week & 6.42  & 6.71  & 0.29  & 0 \\
		Average number of delayed hours for each delayed trip & 0.77  & 0.77  & 0.00  & 0.3947 \\
		Average number of delayed hours for all trips & 0.18  & 0.19  & 0.01  & 0 \\
		\bottomrule
	\end{tabular}}
	\label{tbl_2city_buff_method2}%
\end{table}%

\indent From Table \ref{tbl_2city_buff_method2}, we can draw conclusions similar to those obtained from the Set-covering-type Model. One difference is the percentage of trips driven from the home base of the engineers. In the Direct Algorithm results, a higher upper bound on the number of engineers leads to a significantly higher percentage. This is because when there is sufficient manpower in the queue  at the home base waiting to be assigned, it is unnecessary to have engineers drive from their away base back to their home base.

\indent In addition to the change in cost metrics, the change in the impact of call window lengths, the maximum number of delayed hours, and whether to allow engineers to drive a train home are all insignificant. In brief, the impacts of those parameters on the cost metrics are the same for $\beta=1.0$ and $\beta=1.2$.

\subsubsection{Impact of $\beta$ on 3-city Instances}

\noindent The sensitivity analysis of $\beta$ on 3-city instances is given in Appendix \ref{appendix_results_discussion_method2_3city}.

\section{Comparison of Results from Set-covering-type Model and Direct Algorithm} 
\label{sec_compexp_compare}

\noindent  Table \ref{tbl_comparison} summarizes the differences in output obtained from the two models for the 2-city and 3-city instances with $\beta=1.2$. Values in columns 2, 3, 5 and 6 are the average of the cost metrics obtained from all 600 cases with different call widow lengths, maximum number of delayed hours, and whether to force engineers to return home by van, where $600=(25 \text{ instances}) \times (4 \text{ call window lengths}) \times (3 \text{ maximum number of delayed hours}) \times (2 \text{ options for returning home})$.  The p-values calculated from the paired t-tests are given in columns 4 and 7, respectively.


\begin{table}[htbp]
	\centering
	\caption{Comparison of costs obtained from Set-covering-type Model and Direct Algorithm for $\beta=1.2$}
	\resizebox{\columnwidth}{!}{\begin{tabular}{|l|c|c|c|c|c|c|}
		\toprule
		\multicolumn{1}{|l|}{\multirow{2}[4]{*}{Cost}} & \multicolumn{3}{c|}{2-city instances} & \multicolumn{3}{c|}{3-city instances} \\
		\cmidrule{2-7}          & \multicolumn{1}{p{4.145em}|}{Set-cover} & \multicolumn{1}{p{4.97em}|}{Direct Alg.} & p-value  & \multicolumn{1}{p{4.145em}|}{Set-cover} & \multicolumn{1}{p{4.97em}|}{Direct Alg.} & p-value \\
		\midrule
		Avg. undercoverage rate (\%) & 5.11  & 6.04  & 0     & 8.91  & 28.17 & 0 \\
		Avg. no. working engineers & 18.02 & 17.90 & 0 & 35.23 & 26.62 & 0 \\
		Avg. no. engineers with at least one 48-hr rest in a row per week & 16.40 & 15.99 & 0     & 33.06 & 24.85 & 0 \\
		Avg. no. train trips for each engineer per week & 2.22  & 2.21  & 0.0194  & 2.46  & 2.57  & 0 \\
		Avg. no. driving hours for each engineer per week & 18.45 & 18.39 & 0.0727 & 23.50 & 24.54 & 0 \\
		Avg. no. working hours for each engineer per week & 25.12 & 25.04 & 0.0548 & 30.89 & 32.24 & 0 \\
		Percentage of trips driven by engineers from home base (\%) & 96.37 & 98.99 & 0     & 96.95 & 99.05 & 0 \\
		Avg. no. delayed trips per week & 2.66  & 6.71  & 0     & 9.65  & 14.74 & 0 \\
		Avg. no. delayed hours for each delayed trip & 0.84  & 0.77  & 0     & 0.84  & 0.79  & 0 \\
		Avg. no. delayed hours for all trips & 0.08  & 0.19  & 0     & 0.14  & 0.19  & 0 \\
		\bottomrule
	\end{tabular}}
	\label{tbl_comparison}%
\end{table}%

\indent From the table, it can be seen that (i) the Set-covering-type Model provides a significantly lower undercoverage rate than the Direct Algorithm in both the 2-city and 3-city instances; (ii) for both 2-city and 3-city instances, the solutions from the Direct Algorithm have significantly fewer working engineers than those from the Set-covering-type Model; (iii) the Set-covering-type Model produces call window schedules with a significantly greater number of engineers having at least one consecutive 48-hour rest  period in a week; (iv) in the 2-city instances, the Set-covering-type Model assigns a slightly higher workload to each engineer, while in the 3-city instances, the Direct Algorithm assigns a significantly higher workload; (v) the percentage of trips driven by engineers from their home base is significantly lower for the Set-covering-type Model; (vi) in both the 2-city and 3-city instances, the Set-covering-type Model provides significantly fewer delayed trips and a smaller number of delayed hours than the Direct Algorithm. In summary, the Set-covering-type Model outperforms the Direct Algorithm with respect to demand coverage, consecutive 48-hour rest, and trip delay.

\section{Discussion and Conclusions} \label{sec_conclusion}

\noindent This study addressed the call window scheduling problem for freight rail drivers under the Hours of Service regulations in the U.S. We proposed and compared two methodologies — a Set-covering-type optimization model and a Direct Algorithm — to determine call window schedules that maximize demand coverage while minimizing the number of required engineers. Computational experiments on 2-city and 3-city instances with varying data and parameter settings revealed notable differences in the performance and operational characteristics of the two approaches.

\indent The Set-covering-type model produced significantly lower demand undercoverage (5.11\% for 2-city and 8.91\% for 3-city instances) compared with the Direct Algorithm. It also generated schedules with a higher probability of engineers obtaining consecutive 48-hour rest periods as well as reduced start time delays. These findings suggest that the Set-covering-type model promotes a healthier and more sustainable work environment for drivers, with potential benefits for long-term workforce satisfaction and safety. For both approaches, the upper limit on the number of available engineers emerged as a critical factor, significantly influencing demand coverage, consecutive rest periods, total workload, and delays.

\indent Nevertheless, the work has several limitations. The analysis assumed deterministic demand scenarios and relied on discrete event simulation to address uncertainty in inter-departure times and trip lengths.  Although the schedules appear to be robust in light of the disruptions modeled, the approach may not fully capture the stochastic nature of freight demand. Future research could extend this study by investigating other stochastic demand models. In addition, investigating the impacts of driver preferences, union agreements, and multi-day duty patterns could yield richer and more practical scheduling frameworks for real-world freight operations. Finally, we could incorporate engineer routing and qualification constraints in future studies by extending the model to account for engineer qualifications, equipment types, and regional operating rules. This might better reflect operational realities where not all engineers are interchangeable, and could reveal new trade-offs between flexibility, demand coverage, and compliance. \\
\\
\\
\noindent \textbf{\Large{References}} \\

\noindent Andrade-Michel, A., Y.A. Ríos-Solís and V. Boyer (2021). Vehicle and Reliable Drive Scheduling for Public Bus Transport Systems. \textit{Transportation Research Part B} 145, 290-301. \\[-5pt]

\noindent Boyer, V., O.J. Ibarra-Rojas and Y.Á. Ríos-Solís (2018). Vehicle and Crew Scheduling for Flexible Bus Transportation Systems. \textit{Transportation Research Part B} 112, 216-229. \\[-5pt]

\noindent El-Rifai, O., T. Garaix and X. Xie (2016). Proactive On-call Scheduling during a Seasonal Epidemic. \textit{Operations Research for Health Care} 8, 53-61. \\[-5pt]

\noindent Feng, T., R.M. Lusby, Y. Zhang, S. Tao, B. Zhang and Q. Peng (2024). A Branch-and-price Algorithm for Integrating Urban Rail Crew Scheduling and Rostering Problems. \textit{Transportation Research Part B} 183, 102941. \\
DOI: https://doi.org/10.1016/j.trb.2024.102941 \\[-5pt]

\noindent Feng, T., R.M. Lusby, Y. Zhang, Q. Peng, P. Shang and S. Tao (2023). An ADMM-based Dual Decomposition Mechanism for Integrating Crew Scheduling and Rostering in an Urban Rail Transit Line. \textit{Transportation Research Part C} 149, 104081. \\
DOI: https://doi.org/10.1016/j.trc.2023.104081 \\[-5pt]

\noindent Frisch, S., P. Hungerländer, A. Jellen (2022). On a Real-World Railway Crew Scheduling Problem. \textit{Transportation Research Procedia} 62, 824-831. \\[-5pt]

\noindent Fuentes, M., L. Cadarso and Á. Marín (2019).  A Hybrid Model for Crew Scheduling in Rail Rapid Transit Networks. \textit{Transportation Research Part B} 125, 248-265.  \\[-5pt]

\noindent Gattermann-Itschert, T., L.M. Poreschack, U.W. Thonemann (2023). Using Machine Learning to Include Planners’ Preferences in Railway Crew Scheduling Optimization. \textit{Transportation Science} 57(3), 796-812. \\[-5pt]

\noindent Gawas, P., A. Legrain and L. Rousseau (2023). Notification Timing for On-Demand Personnel Scheduling. \textit{Production and Operations Management} Dec., 2023. \\[-5pt]

\noindent Grover, L.K. (1996). A Fast Quantum Mechanical Algorithm for Database Search. \textit{Proceedings of the Twenty-eighth Annual ACM Symposium on Theory of Computing} 212-219. ACM, New York. \\[-5pt]

\noindent Guo, J. and J.F. Bard (2024). Weekly Scheduling for Freight Rail Engineers \& Trainmen. \textit{Transportation Research Part B} 183, 102942. \\
DOI: https://doi.org/10.1016/j.trb.2024.102942 \\[-5pt]

\noindent Hanafi, R. and E. Kozan (2014). A Hybrid Constructive Heuristic and Simulated Annealing for Railway Crew Scheduling. \textit{Computers \rm{\&} Industrial Engineering} 70, 11-19. \\[-5pt]

\noindent Heil, J., K. Hoffmann and U. Buscher (2020). Railway Crew Scheduling: Models, Methods and Applications. \textit{European Journal of Operational Research} 283, 405-425. \\[-5pt]

\noindent Janacek, J., M. Kohani, M. Koniorczyk and P. Marton (2017). Optimization of Periodic Crew Schedules with Application of Column Generation Method. \textit{Transportation Research Part C} 83, 165-178. \\[-5pt]

\noindent Jütte, S. and U.W. Thonemann (2012). Divide-and-price: A Decomposition Algorithm for Solving Large Railway Crew Scheduling Problems. \textit{European Journal of Operational Research} 219(2), 214-223. \\[-5pt]

\noindent Jütte, S, D. Müller and U.W. Thonemann (2017). Optimizing Railway Crew Schedules with Fairness Preferences. \textit{Journal of Scheduling} 20, 43-55. \\[-5pt]

\noindent Jütte, S., M. Albers, U.W. Thonemann and K. Haase (2011). Optimizing Railway Crew Scheduling at DB Schenker. \textit{Interfaces} 41(2), 109-122. \\[-5pt]

\noindent Lyu, J. and J.F. Bard (2025). Weekly Crew Scheduling for Freight Rail Engineers: A Network Approach. \textit{Journal of Rail Transport Planning \& Management} 34, 100519.\\
DOI: https://doi.org/10.1016/j.jrtpm.2025.100519  \\[-5pt]

\noindent Nishi, T., Y. Muroi and M. Inuiguchi (2011). Column Generation with Dual Inequalities for Railway Crew Scheduling Problems. \textit{Public Transport} 3, 25-42.  \\[-5pt]

\noindent Public Law 110-432 (2008). Federal Rail Safety Improvements. \\
https://www.congress.gov/110/plaws/publ432/PLAW-110publ432.pdf \\[-5pt]

\noindent Rählmann, C., F. Wagener, U.W. Thonemann (2021). Robust Tactical Crew Scheduling Under Uncertain Demand. \textit{Transportation Science} 55(6), 1392-1410. \\[-5pt]

\noindent Rählmann, C., U.W. Thonemann (2020). Railway Crew Scheduling with Semi-flexible Timetables. \textit{OR Spectrum} 42, 835-862.  \\[-5pt]

\noindent Scherer, A., T. Guggemos, S. Grundner-Culemann, N. Pomplun, S. Prüfer and A. Spörl (2021). OnCall Operator Scheduling for Satellites with Grover’s Algorithm. \textit{Computational Science - ICCS} 2021, 12747, 17-29. \\[-5pt]

\noindent Shamia, O., N. Aboushaqrah and N. Bayoumy (2015). Physician On Call Scheduling: Case of a Qatari Hospital. \textit{2015 6th International Conference on Modeling, Simulation, and Applied Optimization (ICMSAO)}, 1-6. \\[-5pt]

\noindent Shen, Y., K. Peng, K. Chen and J. Li (2013). Evolutionary Crew Scheduling with Adaptive Chromosomes. \textit{Transportation Research Part B} 56, 174-185. \\[-5pt]

\noindent Vaidyanathan, B., K.C. Jha and R.K. Ahuja (2007). Multicommodity Network Flow Approach to the Railroad Crew-Scheduling Problem. \textit{IBM Journal of Research and Development} 51(3/4), 325-344.  \\[-5pt]

\noindent Van Rossum, B., T. Dollevoet and D. Hulsman (2025). Benders Decomposition for Robust Tactical Railway Crew Scheduling. \textit{Transportation Science} 59(6), 1283-1302.  \\[-5pt]

\noindent Wang, M., S. Chen and Q. Meng (2024). Robust Safety Driver Scheduling for Autonomous Buses. \textit{Transportation Research Part B} 184, 102965. \\
DOI: https://doi.org/10.1016/j.trb.2024.102965  \\[-5pt]

\noindent Xu, Y., H. Yin, S. Yang, H. Zheng, X. Chang and J. Wu (2025). Cross-line Crew Scheduling Optimization in Urban Rail Transit Systems. \textit{Computers \& Industrial Engineering} 201, 110896. \\
DOI: https://doi.org/10.1016/j.cie.2025.110896 \\[-5pt]

\newpage

\begin{appendices}
	
		\section{Tablularized result for 2-city and 3-city instances obtained from Method 1: Set-covering-type Model with $\beta=1.2$} \label{appendix_results_2/3city}

\noindent Table \ref{tbl_2city_summary} summarizes the 2-city cost metrics obtained from the Set-covering-type Model for all scenarios with $\beta=1.2$ , i.e., combinations of different call window lengths, maximum number of delayed hours, and whether to require engineers to return home by van. Columns 1-3 list the call window length, maximum number of delayed hours, and whether to require engineers to return home by van, respectively. Columns 4-11 provide the average cost metrics. The last column gives the average simulation times. 

With $\beta=1.2$, we ran 25 2-city instances, so each row in Table \ref{tbl_2city_summary} is the average value for 25 instances. For example, in row 2 column 4, the value is 14.10. This indicates in the scenario where the call window length is 6 hours, the maximum number of delayed hours is 0, and engineers are always required to return home by van, that the average undercoverage rate is 14.10\%. 

Tables \ref{tbl_2city_length}-\ref{tbl_2city_YN} display the means and statistical test p-values for 2-city costs as a function of call window lengths, maximum delayed hours, and whether to requrie engineers to return home by van, respectively. In particular, in Table \ref{tbl_2city_length}, column 1 identifies the costs. Columns 2-5 show the cost values when the call window lengths are 6, 10, 12 and 24 hours, respectively. Each value is the average of 150 cases obtained from 25 instances, three possible maximum delayed hours, and two options for returning home. Column 6 is the p-values from the ANOVA test. The last column summarizes whether the call window length significantly impacts the cost metric.

Note that each value in Tables \ref{tbl_2city_length}-\ref{tbl_2city_YN} is obtained by calculating the corresponding average value from Table \ref{tbl_2city_summary}. For example, the value in Table \ref{tbl_2city_length} for row 2, column 2 is 10.85, indicating that the average undercoverage rate for all scenarios with 6-hour call windows is 10.85\%. This value is obtained by calculating the average of the undercoverage rates in rows 2-7, column 4 of Table \ref{tbl_2city_summary}, i.e., $(14.10+11.95+11.47+9.91+9.52+8.13)/6 \approx 10.85$.

Tables \ref{tbl_3city_summary}-\ref{tbl_3city_YN} follow the similar logic of Tables \ref{tbl_2city_summary}-\ref{tbl_2city_YN} but focus on 3-city instances. Table \ref{tbl_3city_summary} summarizes the 3-city cost metrics obtained from the Set-covering-type Model for all scenarios with $\beta=1.2$. Tables \ref{tbl_3city_length}-\ref{tbl_3city_YN} display the means and statistical test p-values of the 3-city costs as a function of call window lengths, maximum delayed hours, and whether to require engineers to return home by van, respectively.

\begin{landscape}
	\begin{table}[htbp]
		\centering
		\caption{Average result statistics for 25 2-city instances obtained from Set-covering-type Model for all scenarios with $\beta=1.2$ }
		\resizebox{\columnwidth}{!}{
			\begin{tabular}{|c|c|c|c|c|c|c|c|c|c|c|c|c|c|c|}
				\toprule
				\multicolumn{1}{|p{4.125em}|}{Call window length (hr)} & \multicolumn{1}{p{4.125em}|}{Maximum delay (hr)} & \multicolumn{1}{p{4.3em}|}{Always take van home} & \multicolumn{1}{p{4.3em}|}{Avg. undercoverage rate (\%)} & \multicolumn{1}{p{4.3em}|}{Avg. no. working engineers} & \multicolumn{1}{p{4.3em}|}{Avg. no. engineers with at least one 48-hr rest in a row per week} & \multicolumn{1}{p{4.3em}|}{Avg. no. train trips for each engineer per week} & \multicolumn{1}{p{4.3em}|}{Avg. no. driving hours for each engineer per week} & \multicolumn{1}{p{4.3em}|}{Avg. no. working hours for each engineer per week} & \multicolumn{1}{p{4.3em}|}{Avg. no. trips driven by engineers from home base} & \multicolumn{1}{p{4.3em}|}{Avg. no. trips driven by engineers from away base} & \multicolumn{1}{p{4.3em}|}{Avg. no. delayed trips per week} & \multicolumn{1}{p{4.3em}|}{Avg. no. delayed hours for each delayed trip} & \multicolumn{1}{p{4.3em}|}{Avg. no. delayed hours for all trips} & \multicolumn{1}{p{4.3em}|}{Simulation runtime (sec)} \\
				\midrule
				\multirow{6}[6]{*}{6} & \multirow{2}[2]{*}{0} & Y     & 14.10 & 18.12 &  16.96 & 2.00  & 16.57 & 22.57 & 144.40 & 0.00  & 0.00  & 0.00  & 0.00  & 312.34 \\
				&       & N     & 11.95 & 18.12 &   16.86 & 2.05  & 17.06 & 23.21 & 132.72 & 15.32 & 0.00  & 0.00  & 0.00  & 267.26 \\
				\cmidrule{2-15}          & \multirow{2}[2]{*}{1} & Y     & 11.47 & 18.12 &   16.98 & 2.06  & 17.08 & 23.25 & 148.80 & 0.00  & 3.10  & 1.00  & 0.07  & 257.42 \\
				&       & N     & 9.91  &  18.12  & 16.73 & 2.10  & 17.39 & 23.68 & 136.08 & 15.36 & 2.69  & 1.00  & 0.06  & 260.55 \\
				\cmidrule{2-15}          & \multirow{2}[2]{*}{2} & Y     & 9.52  & 18.12 &   16.95 & 2.11  & 17.45 & 23.76 & 152.12 & 0.00  & 5.82  & 1.49  & 0.21  & 267.81 \\
				&       & N     & 8.13  & 18.12 &   16.73 & 2.14  & 17.73 & 24.15 & 139.32 & 15.16 & 4.97  & 1.49  & 0.18  & 266.44 \\
				\midrule
				\multirow{6}[6]{*}{10} & \multirow{2}[2]{*}{0} & Y     & 7.11  & 18.12 &   17.05 & 2.16  & 17.95 & 24.44 & 156.08 & 0.00  & 0.00  & 0.00  & 0.00  & 379.07 \\
				&       & N     & 5.90  & 18.12   & 16.61 & 2.19  & 18.19 & 24.77 & 142.56 & 15.60 & 0.00  & 0.00  & 0.00  & 400.65 \\
				\cmidrule{2-15}          & \multirow{2}[2]{*}{1} & Y     & 5.98  & 18.12 &   16.82 & 2.19  & 18.18 & 24.75 & 158.00 & 0.00  & 3.02  & 1.00  & 0.07  & 385.72 \\
				&       & N     & 4.75  & 18.12 &   16.56 & 2.22  & 18.46 & 25.12 & 144.28 & 15.80 & 2.63  & 1.00  & 0.06  & 405.24 \\
				\cmidrule{2-15}          & \multirow{2}[2]{*}{2} & Y     & 4.64  & 18.12 &   16.61 & 2.22  & 18.44 & 25.10 & 160.28 & 0.00  & 5.57  & 1.51  & 0.20  & 411.61 \\
				&       & N     & 3.31  & 18.12 &   16.38 & 2.25  & 18.72 & 25.48 & 146.80 & 15.72 & 5.41  & 1.52  & 0.19  & 445.20 \\
				\midrule
				\multirow{6}[6]{*}{12} & \multirow{2}[2]{*}{0} & Y     & 4.66  & 18.12 &   16.74 & 2.22  & 18.43 & 25.10 & 160.20 & 0.00  & 0.00  & 0.00  & 0.00  & 470.91 \\
				&       & N     & 3.63  & 18.12 &   16.46 & 2.25  & 18.66 & 25.40 & 147.52 & 14.36 & 0.00  & 0.00  & 0.00  & 464.15 \\
				\cmidrule{2-15}          & \multirow{2}[2]{*}{1} & Y     & 3.64  & 18.12 &   16.64 & 2.25  & 18.65 & 25.38 & 161.88 & 0.00  & 3.12  & 1.00  & 0.07  & 488.48 \\
				&       & N     & 3.07  & 18.04 &   16.36 & 2.27  & 18.83 & 25.63 & 148.48 & 14.36 & 2.87  & 1.00  & 0.07  & 490.54 \\
				\cmidrule{2-15}          & \multirow{2}[2]{*}{2} & Y     & 3.44  & 17.88 &   16.28 & 2.28  & 18.97 & 25.82 & 162.28 & 0.00  & 5.60  & 1.52  & 0.20  & 524.45 \\
				&       & N     & 2.26    & 18.00 & 16.26 & 2.29  & 19.03 & 25.90 & 149.60 & 14.60 & 5.37  & 1.50  & 0.19  & 478.45 \\
				\midrule
				\multirow{6}[6]{*}{24} & \multirow{2}[2]{*}{0} & Y     & 0.93  & 17.92 &   15.61 & 2.34  & 19.36 & 26.36 & 166.48 & 0.00  & 0.00  & 0.00  & 0.00  & 918.93 \\
				&       & N     & 1.15  & 17.84 &   15.73 & 2.34  & 19.40 & 26.41 & 165.24 & 0.84  & 0.00  & 0.00  & 0.00  & 912.88 \\
				\cmidrule{2-15}          & \multirow{2}[2]{*}{1} & Y     & 0.98  & 17.80 &   15.63 & 2.35  & 19.51 & 26.56 & 166.36 & 0.00  & 2.27  & 1.00  & 0.05  & 862.14 \\
				&       & N     & 0.77  & 17.64 &   15.61 & 2.38  & 19.70 & 26.83 & 165.76 & 0.96  & 2.30  & 1.00  & 0.05  & 868.17 \\
				\cmidrule{2-15}          & \multirow{2}[2]{*}{2} & Y     & 0.70  & 17.76 &   15.44 & 2.36  & 19.58 & 26.66 & 166.84 & 0.00  & 4.63  & 1.52  & 0.17  & 956.09 \\
				&       & N     & 0.63  & 17.80 &   15.49 & 2.36  & 19.55 & 26.62 & 166.04 & 0.92  & 4.54  & 1.52  & 0.16  & 940.98 \\
				\bottomrule
		\end{tabular}}
		\label{tbl_2city_summary}%
	\end{table}%

	\begin{table}[htbp]
		\centering
		\caption{Means and ANOVA p-values of costs with each call window length for 2-city instances obtained from the Set-covering-type Model for all scenarios with $\beta=1.2$ }
		\resizebox{\columnwidth}{!}{\begin{tabular}{|l|c|c|c|c|c|c|}
				\hline
				\multicolumn{1}{|p{4.125em}|}{Cost}  & \multicolumn{1}{p{4.125em}|}{6-hr} & \multicolumn{1}{p{4.125em}|}{10-hr} & \multicolumn{1}{p{4.125em}|}{12-hr} & \multicolumn{1}{p{4.125em}|}{24-hr} & \multicolumn{1}{p{4.125em}|}{p-value} & \multicolumn{1}{p{4.125em}|}{Significant impact} \\ \hline
				Undercoverage rate (\%) & 10.85 & 5.28  & 3.45  & 0.86  & 0 & \checkmark \\
				Average number of working engineers & 18.12 & 18.12 & 18.05 & 17.79 & 0.3583 & \\
				Average number of engineers with at least one 48-hr rest in a row per week & 16.87 & 16.67 & 16.46 & 15.59 & 0 & \checkmark \\
				Average number of train trips for each engineer per week & 2.07  & 2.21  & 2.26  & 2.35  & 0 & \checkmark \\
				Average number of driving hours for each engineer per week & 17.21 & 18.32 & 18.76 & 19.52 & 0 & \checkmark \\
				Average number of working hours for each engineer per week & 23.44 & 24.94 & 25.54 & 26.57 & 0 & \checkmark \\
				Percentage of trips driven from engineer's home base (\%) & 94.96 & 95.11 & 95.56 & 99.73 & 0 & \checkmark \\
				Average number of delayed trips per week & 2.76  & 2.77  & 2.83  & 2.29  & 0.1654 &  \\
				Average number of delayed hours for each delayed trip & 0.83  & 0.84  & 0.84  & 0.84  & 0.9992 &  \\
				Average number of delayed hours for all trips & 0.09  & 0.09  & 0.09  & 0.07  & 0.3453 &  \\ \hline
		\end{tabular}}
		\label{tbl_2city_length}%
	\end{table}%

	\begin{table}[htbp]
		\centering
		\caption{Means and ANOVA p-values of costs with each upper limit on the number of delayed hours for 2-city instances obtained from Set-covering-type Model for all scenarios with $\beta=1.2$ }
		\resizebox{\columnwidth}{!}{\begin{tabular}{|l|c|c|c|c|c|}
				\hline
				Cost  & \multicolumn{1}{p{4.125em}|}{0-hr} & \multicolumn{1}{p{4.125em}|}{1-hr} & \multicolumn{1}{p{4.125em}|}{2-hr} & \multicolumn{1}{p{4.125em}|}{p-value} & \multicolumn{1}{p{4.125em}|}{Significant impact} \\ \hline
				Undercoverage rate (\%) & 6.18  & 5.07  & 4.08  & 0 & \checkmark \\
				Average number of working engineers & 18.06 & 18.01 & 17.99 & 0.9255 & \\
				Average number of engineers with at least one 48-hr rest in a row per week & 16.50 & 16.42 & 16.27 & 0.2767 &  \\
				Average number of train trips for each engineer per week & 2.19  & 2.23  & 2.25  & 0.0051 & \checkmark \\
				Average number of driving hours for each engineer per week & 18.20 & 18.47 & 18.68 & 0.0091 & \checkmark \\
				Average number of working hours for each engineer per week & 24.78 & 25.15 & 25.44 & 0.0076 & \checkmark \\
				Percentage of trips driven from engineer's home base (\%) & 96.30 & 96.33 & 96.39 & 0.9841 &  \\
				Average number of delayed trips per week & 0.00  & 2.75  & 5.24  & 0 & \checkmark \\
				Average number of delayed hours for each delayed trip & 0.00  & 1.00  & 1.51  & 0 & \checkmark \\
				Average number of delayed hours for all trips & 0.00  & 0.07  & 0.19  & 0 & \checkmark \\ \hline
		\end{tabular}}
		\label{tbl_2city_delayhour}%
	\end{table}%

	\begin{table}[htbp]
		\centering
		\caption{Means and paired t-test p-values of costs under conditions whether to always require engineers to return home by van (Y), or to allow engineers to drive a train home (N) for 2-city instances obtained from Set-covering-type Model for all scenarios with $\beta=1.2$}
		\resizebox{\columnwidth}{!}{\begin{tabular}{|l|c|c|c|c|}
				\hline
				Cost  & \multicolumn{1}{p{4.125em}|}{Y (van home)} & \multicolumn{1}{p{4.125em}|}{N (allow driving home)} & \multicolumn{1}{p{4.125em}|}{p-value} & \multicolumn{1}{p{4.125em}|}{Significant impact} \\ \hline
				Undercoverage rate (\%) & 5.60  & 4.62  & 0 & \checkmark \\  
				Average number of working engineers & 18.01 & 18.03 & 0.6064 & \\
				Average number of engineers with at least one 48-hr rest in a row per week & 16.48 & 16.32 & 0 & \checkmark \\
				Average number of train trips for each engineer per week & 2.21  & 2.24  & 0 & \checkmark \\
				Average number of driving hours for each engineer per week & 18.35 & 18.56 & 0 & \checkmark \\
				Average number of working hours for each engineer per week & 24.98 & 25.27 & 0 & \checkmark \\
				Percentage of trips driven from engineer's home base (\%) & 100.00 & 92.68 & 0 & \checkmark \\
				Average number of delayed trips per week & 2.76  & 2.57  & 0 & \checkmark \\
				Average number of delayed hours for each delayed trip & 0.84  & 0.84  & 0.4797 &  \\
				Average number of delayed hours for all trips & 0.09  & 0.08  & 0 & \checkmark \\ \hline
		\end{tabular}}
		\label{tbl_2city_YN}%
	\end{table}%

	\begin{table}[htbp]
		\centering
		\caption{Average results for 25 3-city instances obtained from Set-covering-type Model for all scenarios with $\beta=1.2$}
		\resizebox{\columnwidth}{!}{
			\begin{tabular}{|c|c|c|c|c|c|c|c|c|c|c|c|c|c|c|}
				\toprule
				\multicolumn{1}{|p{4.125em}|}{Call window length (hr)} & \multicolumn{1}{p{4.125em}|}{Maximum delay (hr)} & \multicolumn{1}{p{4.3em}|}{Always take van home} & \multicolumn{1}{p{4.3em}|}{Avg. undercoverage rate (\%)} & \multicolumn{1}{p{4.3em}|}{Avg. no. working engineers} & \multicolumn{1}{p{4.3em}|}{Avg. no. engineers with at least one 48-hr rest in a row per week} & \multicolumn{1}{p{4.3em}|}{Avg. no. train trips for each engineer per week} & \multicolumn{1}{p{4.3em}|}{Avg. no. driving hours for each engineer per week} & \multicolumn{1}{p{4.3em}|}{Avg. no. working hours for each engineer per week} & \multicolumn{1}{p{4.3em}|}{Avg. no. trips driven by engineers from home base} & \multicolumn{1}{p{4.3em}|}{Avg. no. trips driven by engineers from away base} & \multicolumn{1}{p{4.3em}|}{Avg. no. delayed trips per week} & \multicolumn{1}{p{4.3em}|}{Avg. no. delayed hours for each delayed trip} & \multicolumn{1}{p{4.3em}|}{Avg. no. delayed hours for all trips} & \multicolumn{1}{p{4.3em}|}{Simulation runtime (sec)} \\
				\midrule
				\multirow{6}[6]{*}{6} & \multirow{2}[2]{*}{0} & Y     & 19.19 & 35.24  & 33.49 & 2.18  & 20.73 & 27.28 & 307.12 & 0.00  & 0.00  & 0.00  & 0.00  & 448.99 \\
				&       & N     & 16.83 & 35.24  & 33.06 & 2.25  & 21.39 & 28.13 & 287.28 & 28.88 & 0.00  & 0.00  & 0.00  & 414.92 \\
				\cmidrule{2-15}          & \multirow{2}[2]{*}{1} & Y     & 16.69 &  35.24  & 33.43 & 2.25  & 21.36 & 28.11 & 316.52 & 0.00  & 11.32 & 1.00  & 0.12  & 422.68 \\
				&       & N     & 14.33 &  35.24  & 32.93 & 2.31  & 22.03 & 28.98 & 296.04 & 29.52 & 10.51 & 1.00  & 0.11  & 424.06 \\
				\cmidrule{2-15}          & \multirow{2}[2]{*}{2} & Y     & 14.17 &  35.24  & 33.55 & 2.32  & 22.07 & 29.03 & 326.16 & 0.00  & 20.75 & 1.53  & 0.33  & 422.26 \\
				&       & N     & 12.30 & 35.24  & 33.02 & 2.37  & 22.57 & 29.68 & 303.96 & 29.36 & 18.80 & 1.53  & 0.30  & 432.73 \\
				\midrule
				\multirow{6}[6]{*}{10} & \multirow{2}[2]{*}{0} & Y     & 11.90 &  35.24 & 33.70 & 2.38  & 22.71 & 29.85 & 334.80 & 0.00  & 0.00  & 0.00  & 0.00  & 669.47 \\
				&       & N     & 10.23 & 35.24  & 33.50 & 2.43  & 23.16 & 30.44 & 313.56 & 27.52 & 0.00  & 0.00  & 0.00  & 669.96 \\
				\cmidrule{2-15}          & \multirow{2}[2]{*}{1} & Y     & 9.70 &   35.24  & 33.71 & 2.44  & 23.30 & 30.61 & 343.04 & 0.00  & 11.17 & 1.00  & 0.12  & 676.20 \\
				&       & N     & 9.22 &  35.12  & 33.16 & 2.46  & 23.45 & 30.83 & 316.60 & 28.08 & 10.10 & 1.00  & 0.11  & 674.07 \\
				\cmidrule{2-15}          & \multirow{2}[2]{*}{2} & Y     & 8.75 &  35.24   & 33.46 & 2.47  & 23.54 & 30.94 & 346.76 & 0.00  & 19.85 & 1.52  & 0.32  & 690.18 \\
				&       & N     & 7.60 &  35.65  & 33.37 & 2.50  & 23.89 & 31.40 & 328.53 & 27.41 & 19.09 & 1.54  & 0.30  & 713.10 \\
				\midrule
				\multirow{6}[6]{*}{12} & \multirow{2}[2]{*}{0} & Y     & 9.63 &  35.24  & 33.53 & 2.44  & 23.32 & 30.65 & 343.48 & 0.00  & 0.00  & 0.00  & 0.00  & 768.42 \\
				&       & N     & 8.22 &  35.24  & 33.20 & 2.48  & 23.69 & 31.13 & 321.68 & 27.12 & 0.00  & 0.00  & 0.00  & 778.45 \\
				\cmidrule{2-15}          & \multirow{2}[2]{*}{1} & Y     & 8.02  &  35.24  & 33.48 & 2.49  & 23.75 & 31.21 & 349.64 & 0.00  & 11.30 & 1.00  & 0.12  & 773.08 \\
				&       & N     & 6.90  & 35.24  & 33.04 & 2.52  & 24.03 & 31.58 & 326.92 & 26.96 & 10.01 & 1.00  & 0.10  & 798.64 \\
				\cmidrule{2-15}          & \multirow{2}[2]{*}{2} & Y     & 7.34 &  35.24   & 33.26 & 2.51  & 23.93 & 31.45 & 352.24 & 0.00  & 19.28 & 1.53  & 0.31  & 793.21 \\
				&       & N     & 5.93 &  35.24  & 33.07 & 2.54  & 24.31 & 31.94 & 331.52 & 26.04 & 18.74 & 1.53  & 0.30  & 795.51 \\
				\midrule
				\multirow{6}[6]{*}{24} & \multirow{2}[2]{*}{0} & Y     & 3.36 &  35.24  & 32.50 & 2.62  & 24.98 & 32.83 & 367.36 & 0.00  & 0.00  & 0.00  & 0.00  & 1544.75 \\
				&       & N     & 3.55 &  35.24  & 32.54 & 2.61  & 24.94 & 32.77 & 365.68 & 1.04  & 0.00  & 0.00  & 0.00  & 1545.89 \\
				\cmidrule{2-15}          & \multirow{2}[2]{*}{1} & Y     & 2.90 &  35.16   & 32.28 & 2.63  & 25.16 & 33.05 & 369.24 & 0.00  & 9.13  & 1.00  & 0.10  & 1531.70 \\
				&       & N     & 2.71 &  35.16  & 32.45 & 2.64  & 25.22 & 33.13 & 368.72 & 1.24  & 9.21  & 1.00  & 0.10  & 1528.26 \\
				\cmidrule{2-15}          & \multirow{2}[2]{*}{2} & Y     & 2.32 &  35.20   & 31.98 & 2.65  & 25.29 & 33.23 & 371.48 & 0.00  & 16.74 & 1.53  & 0.27  & 1624.15 \\
				&       & N     & 2.66 &  35.20   & 32.03 & 2.64  & 25.20 & 33.11 & 369.00 & 1.12  & 16.61 & 1.53  & 0.27  & 1606.22 \\
				\bottomrule
		\end{tabular}}
		\label{tbl_3city_summary}
	\end{table}%

	\begin{table}[htbp]
		\centering
		\caption{Means and ANOVA p-values of costs with each call window length for 3-city instances obtained from Set-covering-type Model for all scenarios with $\beta=1.2$}
		\resizebox{\columnwidth}{!}{\begin{tabular}{|l|c|c|c|c|c|c|}
				\toprule
				\multicolumn{1}{|p{4.125em}|}{Cost}  & \multicolumn{1}{p{4.125em}|}{6-hr} & \multicolumn{1}{p{4.125em}|}{10-hr} & \multicolumn{1}{p{4.125em}|}{12-hr} & \multicolumn{1}{p{4.125em}|}{24-hr} & \multicolumn{1}{p{4.125em}|}{p-value} & \multicolumn{1}{p{4.125em}|}{Significant impact} \\
				\midrule
				Undercoverage rate (\%) & 15.58 & 9.48  & 7.67  & 2.92  & 0 & \checkmark \\
				Average number of working engineers & 35.24 & 35.24 & 35.24 & 35.20 & 0.9985 & \\
				Average number of engineers with at least one 48-hr rest in a row per week & 33.25 & 33.44 & 33.26 & 32.30 & 0 &  \\
				Average number of train trips for each engineer per week & 2.28  & 2.45  & 2.50  & 2.63  & 0 & \checkmark \\
				Average number of driving hours for each engineer per week & 21.69 & 23.35 & 23.84 & 25.13 & 0 & \checkmark \\
				Average number of working hours for each engineer per week & 28.53 & 30.69 & 31.33 & 33.02 & 0 & \checkmark \\
				Percentage of trips driven from engineer's home base (\%) & 95.51 & 96.07 & 96.22 & 99.85 & 0 & \checkmark \\
				Average number of delayed trips per week & 10.23 & 9.88  & 9.89  & 8.62  & 0.2971 &  \\
				Average number of delayed hours for each delayed trip & 0.84  & 0.84  & 0.84  & 0.84  & 1.0000 &  \\
				Average number of delayed hours for all trips & 0.14  & 0.14  & 0.14  & 0.12  & 0.4036 &  \\
				\bottomrule
		\end{tabular}}
		\label{tbl_3city_length}%
	\end{table}%

	\begin{table}[htbp]
		\centering
		\caption{Means and ANOVA p-values of costs with each upper limit on the number of delayed hours for 3-city instances obtained from Set-covering-type Model for all scenarios with $\beta=1.2$}
		\resizebox{\columnwidth}{!}{
			\begin{tabular}{|l|c|c|c|c|c|}
				\toprule
				Cost  & \multicolumn{1}{p{4.125em}|}{0-hr} & \multicolumn{1}{p{4.125em}|}{1-hr} & \multicolumn{1}{p{4.125em}|}{2-hr} & \multicolumn{1}{p{4.125em}|}{p-value} & \multicolumn{1}{p{4.125em}|}{Significant impact}  \\
				\midrule
				Undercoverage rate (\%) & 10.36 & 8.77  & 7.60  & 0 & \checkmark \\
				Average number of working engineers & 35.24 & 35.22 & 35.23 & 0.9964 & \\
				Average number of engineers with at least one 48-hr rest in a row per week & 33.19 & 33.08 & 32.92 & 0.3870 &  \\
				Average number of train trips for each engineer per week & 2.42  & 2.47  & 2.50  & 0 & \checkmark \\
				Average number of driving hours for each engineer per week & 23.12 & 23.54 & 23.85 & 0 & \checkmark \\
				Average number of working hours for each engineer per week & 30.38 & 30.94 & 31.35 & 0 & \checkmark \\
				Percentage of trips driven from engineer's home base (\%) & 96.85 & 96.88 & 97.00 & 0.9181 &  \\
				Average number of delayed trips per week & 0.00  & 10.32 & 18.64 & 0 & \checkmark \\
				Average number of delayed hours for each delayed trip & 0.00  & 1.00  & 1.53  & 0 & \checkmark \\
				Average number of delayed hours for all trips & 0.00  & 0.11  & 0.30  & 0 & \checkmark \\
				\bottomrule
		\end{tabular}}
		\label{tbl_3city_delayhour}%
	\end{table}%

	\begin{table}[htbp]
		\centering
		\caption{Means and paired t-test p-values of costs under conditions whether to always require engineers to go home by van (Y), or to allow engineers to drive a train home (N) for 3-city instances obtained from Set-covering-type Model for all scenarios with $\beta=1.2$}
		\resizebox{\columnwidth}{!}{\begin{tabular}{|l|c|c|c|c|}
				\hline
				Cost  & \multicolumn{1}{p{4.125em}|}{Y (van home)} & \multicolumn{1}{p{4.125em}|}{N (allow driving home)} & \multicolumn{1}{p{4.125em}|}{p-value} & \multicolumn{1}{p{4.125em}|}{Significant impact} \\
				\midrule
				Undercoverage rate (\%) & 9.50  & 8.33  & 0 & \checkmark \\
				Average number of working engineers & 35.23 & 35.23 & $\sim 1$ & \\
				Average number of engineers with at least one 48-hr rest in a row per week & 33.20 & 32.93 & 0 & \checkmark \\
				Average number of train trips for each engineer per week & 2.45  & 2.48  & 0 & \checkmark \\
				Average number of driving hours for each engineer per week & 23.35 & 23.66 & 0 & \checkmark \\
				Average number of working hours for each engineer per week & 30.69 & 31.10 & 0 & \checkmark \\
				Percentage of trips driven from engineer's home base (\%) & 100.00 & 93.82 & 0 & \checkmark \\
				Average number of delayed trips per week & 9.96  & 9.35  & 0 & \checkmark \\
				Average number of delayed hours for each delayed trip & 0.84  & 0.84  & 0.7800 &  \\
				Average number of delayed hours for all trips & 0.14  & 0.13  & 0 & \checkmark \\
				\bottomrule
		\end{tabular}}
		\label{tbl_3city_YN}%
	\end{table}%

\end{landscape}

\section{Computational results from the Set-covering-type Model for 3-city instances with $\beta=1.2$} \label{appendix_results_discussion_method1_3city}

\subsection{Discussion of results for 3 cities}

\noindent Table \ref{tbl_3city_summary} in Appendix \ref{appendix_results_2/3city} gives the average statistics obtained from 25 instances with 3 cities.  In each instance, we ran 24 experiments with different combinations of parameters.  The remainder of this subsection discusses how each parameter impacts the cost metrics. \\ [-10pt]

\noindent \textit{\textbf{Impact of the call window length.}} Table \ref{tbl_3city_length} in Appendix \ref{appendix_results_2/3city} presents the means and ANOVA p-values of costs with each call window length. The results indicate that a longer call windows lead to the following.
\begin{enumerate}[label=(\roman*)]
	\item An insignificant impact on the number of  engineers (p-value 0.9985) \\[-20pt]
	\item Unstable impact on the number of engineers with at least one 48-hour consecutive rest per week (even though the p-value is approximately 0, the cost metric does not increase or decrease as the call window length grows) \\[-20pt]
	\item A greater workload per week (a greater number of trips, driving hours and working hours with p-values close to 0) \\[-20pt]
	\item A higher percentage of trips driven by engineers from their home base (p-value of 0) \\[-20pt]
	\item An insignificant change in delay (average number of delayed trips per week, number of delayed hours per delayed trip, and number of delayed hours for all trips, with p-values of 0.2971, 1.0000 and 0.4036, respectively) \\[-20pt]
\end{enumerate}

\noindent \textit{\textbf{Impact of the maximum number of delayed hours.}}  Table \ref{tbl_3city_delayhour} in Appendix \ref{appendix_results_2/3city} presents the means and ANOVA p-values for costs for each upper limit on the number of delayed hours. The results indicate that a higher upper limit on the number of delayed hours has the following implications.
\begin{enumerate}[label=(\roman*)]
	\item A lower undercoverage rate (p-value 0) \\ [-20pt]
	\item An insignificant impact on the number of  engineers (p-value 0.9964) \\ [-20pt]
	\item An insignificant impact on the number of engineers with at least one 48-hour consecutive rest per week (p-value 0.3870) \\ [-20pt]
	\item A greater workload per week (a greater number of trips, driving hours and working hours with p-values of 0) \\ [-20pt]
	\item An insignificant impact on the percentage of trips driven by engineers from their home base (p-value 0.9181) \\ [-20pt]
	\item More serious delays (a larger number of delayed trips per week, average number of delayed hours for each trip, and average number of delayed hours for all trips, with p-values of 0, 0 and 0, respectively) \\ [-20pt]
\end{enumerate}

\noindent \textit{\textbf{Impact of whether to require engineers to return home by van.}} Table \ref{tbl_3city_YN} in Appendix \ref{appendix_results_2/3city} presents the means and paired t-test p-values for costs depending on whether engineers are always required to return home by van  (Y), or whether they are allowed to drive a train to their home base when possible (N). The table entries show that allowing engineers to return home by driving a train gives the following outcomes.
\begin{enumerate}[label=(\roman*)]
	\item A lower undercoverage rate (p-value 0) \\ [-20pt]
	\item Trivial impact on the number of working engineers \\ [-20pt]
	\item Fewer engineers with at least one 48-hour consecutive rest per week (p-value 0) \\ [-20pt]
	\item A greater workload per week (a greater number of trips, driving hours and working hours with all p-values of 0) \\ [-20pt]
	\item A lower percentage of trips driven by engineers from their home base (p-value 0) \\ [-20pt]
	\item Less serious delays (a smaller number of delayed trips per week, and average number of delayed hours for all trips, with p-values of 0), and insignificant impact on the average number of delayed hours for each delayed trip (p-value 0.7800) \\ [-20pt]
\end{enumerate}

\subsection{Impact of $\beta$ on 3-city instances}

\noindent Table \ref{tbl_3city_buff} presents the average costs for $\beta=1.0$ and $1.2$ in the second and third columns, respectively. Each value in these two columns was obtained by calculating the average of all 600 individual cases, where $600=(25 \text{ instances}) \times (4 \text{ call window lengths}) \times (3 \text{ maximum number of delayed hours}) \times (2 \text{ options for returning home})$.
The fourth column reports the difference in means between the two $\beta$ values; i.e., the difference between the costs for $\beta=1.2$ minus those for  $\beta=1.0$. The last column gives the p-value from the paired t-test. A p-value below 0.05 indicates the difference in the means between $\beta=1.0$ and $1.2$ is significant.

\renewcommand{\arraystretch}{1.3}
\begin{table}[htbp]
	\centering
	\caption{Means and difference in costs for 3-city instances with $\beta=1.0$ and $\beta=1.2$ obtained from the Set-covering-type Model.}
	\resizebox{\columnwidth}{!}{\begin{tabular}{|l|c|c|c|c|c|}
			\hline
			\multicolumn{1}{|p{4.125em}|}{Cost}  & \multicolumn{1}{p{4.125em}|}{Mean ($\beta=1.0$)} &
			\multicolumn{1}{p{4.125em}|}{Mean ($\beta=1.2$)} & \multicolumn{1}{p{4.125em}|}{Difference (1.2 - 1.0)} & p-value \\ \hline
			Undercoverage rate (\%) & 17.49 & 8.91 & -8.58 & 0 \\
			Average number of working engineers & 29.84 & 35.23 & 5.39 & 0 \\
			Average number of engineers with at least one 48-hr rest in a row per week & 28.61 & 33.06 & 4.45 & 0 \\
			Average number of train trips for each engineer per week & 2.63 & 2.46 & -0.17 & 0 \\
			Average number of driving hours for each engineer per week & 25.18 & 23.50 & -1.67 & 0 \\
			Average number of working hours for each engineer per week & 33.07 & 30.89 & -2.17 & 0 \\
			Percentage of trips driven from engineer's home base (\%) & 96.54 & 96.91 & 0.37 & 0 \\
			Average number of delayed trips per week & 9.23 & 9.65 & 0.42 & 0 \\
			Average number of delayed hours for each delayed trip & 0.84 & 0.84 & 0.00 & $0.7962$ \\
			Average number of delayed hours for all trips & 0.13 & 0.14 & 0.01 & 0  \\ \hline
	\end{tabular}}
	\label{tbl_3city_buff}%
\end{table}%

\indent  Table \ref{tbl_3city_buff} shows that we can draw similar conclusions  for the  3-city instances that we did for  the 2-city instances. One difference relates to the percentage of trips driven from the home base of the engineers. In the 2-city case, a higher upper bound on the number of engineers has insignificant impact on this percentage while in the  3-city case the percentage is significantly higher. This is because when there is sufficient manpower in the queue waiting to be assigned at the home base, it is unnecessary to let engineers drive from their away base back home.

\indent In addition to the change in cost metrics, the change in the impact of call window lengths, the maximum number of delayed hours, and whether to allow engineers to drive a train home are all insignificant. In other words, the impacts of those parameters on the cost metrics are same for $\beta=1.0$ and $\beta=1.2$.

\section{Computational results from the Direct Algorithm on 3-city instances with $\beta=1.2$} \label{appendix_results_discussion_method2_3city}

\subsection{Discussion of results for 3-cities}

\noindent Table \ref{tbl_3city_summary_method2} in Appendix \ref{appendix_results_2/3city_method2} gives the average output statistics for 25 instances with 3 cities.  In each instance, we again ran 24 experiments with different combinations of parameter values.  The remainder of this subsection discusses how each parameter impacts the cost metrics. \\ [-10pt]

\noindent \textit{\textbf{Impact of  the call window length.}} Table \ref{tbl_3city_length_method2} in Appendix \ref{appendix_results_2/3city_method2} presents the means and ANOVA p-values for costs for each call window length. It can be seen that a longer call window  leads to the following changes.
\begin{enumerate}[label=(\roman*)]
	\item A lower undercoverage rate (p-value 0) \\[-20pt]
	\item Insignificantly fewer engineers (p-value 0.7203) \\[-20pt]
	\item Smaller number of engineers with at least one 48-hour consecutive rest period per week (p-value 0.0221) \\[-20pt]
	\item Greater workload per week (a greater number of trips, driving hours and working hours with p-values of 0) \\[-20pt]
	\item A lower percentage of trips driven by engineers from their home base (p-value 0) \\[-20pt]
	\item A smaller number of delayed trips per week, and smaller number of delayed hours for all trips, with p-values of 0. The number of delayed hours per delayed trip is not significantly impacted (p-value 0.9460). \\[-20pt]
\end{enumerate}

\indent Similar to the results for the 2-city instances, the cost metrics are monotone in call window length only when it is in the range from 6 to 12 hours (not extended to 24 hours), except for the average number of working engineers. This can be explained by the fact that in the experiments, whether we can assign a trip to an engineer depends on both the call window schedule and the HOS constraints. Therefore, cost metric monotonicity is not guaranteed with respect to the call window length.    \\ [-10pt]

\noindent \textit{\textbf{Impact of the maximum number of delayed hours.}}  Table \ref{tbl_3city_delayhour_method2} in Appendix \ref{appendix_results_2/3city_method2} presents the means and ANOVA p-values for costs for each upper limit on the number of delayed hours. The entries show that a higher upper limit on the number of delayed hours leads to the following.
\begin{enumerate}[label=(\roman*)]
	\item A lower undercoverage rate (p-value 0)  \\[-20pt]
	\item Insignificantly fewer  engineers (p-value 0.9645) \\[-20pt]
	\item An insignificant impact on the number of engineers with at least one 48-hour consecutive rest period per week (p-value 0.5203) \\[-20pt]
	\item A higher workload per week (a greater number of trips, driving hours and working hours with p-values of 0) \\[-20pt]
	\item No significant impact on the percentage of trips driven by engineers from their home base (p-value 0.7160) \\[-20pt]
	\item More serious delays (a greater number of delayed trips per week, average number of delayed hours for each trip, and average number of delayed hours for all trips, with p-values of 0) \\[-20pt]
\end{enumerate}

\noindent \textit{\textbf{Impact of whether to require engineers to return home by van.}} Table \ref{tbl_3city_YN_method2} in Appendix \ref{appendix_results_2/3city_method2} presents the means and paired t-test p-values for costs under the policy to always require engineers to return home by van (Y), or to allow them to drive a train home (N). According to the results, we see that allowing engineers to return home by driving a train has the following outcomes.
\begin{enumerate}[label=(\roman*)]
	\item A lower undercoverage rate (p-value 0) \\[-20pt]
	\item An insignificant change in the number of  engineers (p-value 0.7820) \\[-20pt]
	\item An insignificant change in the number of engineers with at least one 48-hour consecutive rest period per week (p-value 0.1354) \\[-20pt]
	\item A greater workload per week (a greater number of trips, driving hours and working hours with p-values of 0) \\[-20pt]
	\item A lower percentage of trips driven by engineers from their home base (p-value 0) \\[-20pt]
	\item A trivial impact on delays (insignificant change in the number of delayed trips per week, average number of delayed hours for each delayed trip, and average number of delayed hours for all trips, with p-values of 0.1118, 0.2402 and 0.2972, respectively) \\[-20pt]
\end{enumerate}

\subsection{Impact of $\beta$ on 3-city instances}

\noindent Table \ref{tbl_3city_buff_method2} presents the average cost for each $\beta$ value used when solving the 3-city instances. From the table, we see that the results are quite similar to those obtained by the Set-covering-type Model. One difference is the average number of delayed hours for each delayed trip. In the Direct Algorithm results, a higher upper bound on the number of engineers leads to significantly fewer delayed hours.

\indent In addition to the change in the costs, the change in the impact of call window lengths, the maximum number of delayed hours, and whether to allow engineers to drive a train back to their home  base is insignificant. In other words, the impacts of those parameters on the cost metrics are the same for $\beta=1.0$ and $\beta=1.2$.

\begin{table}[htbp]
	\centering
	\caption{Means and difference in costs for 3-city instancess with $\beta=1.0$ and $\beta=1.2$ obtained from the Direct Algorithm.}
	\resizebox{\columnwidth}{!}{\begin{tabular}{|l|c|c|c|c|}
			\hline
			\multicolumn{1}{|p{4.125em}|}{Cost}  & \multicolumn{1}{p{4.125em}|}{Mean ($\beta=1.0$)} &
			\multicolumn{1}{p{4.125em}|}{Mean ($\beta=1.2$)} & \multicolumn{1}{p{4.125em}|}{Difference (1.2 - 1.0)} & p-value \\ \hline
			Undercoverage rate (\%) & 35.71 & 28.17 & -7.55 & 0 \\
			Average number of working engineers & 22.52 & 26.62 & 4.10 & 0  \\
			Average number of engineers with at least one  48-hr rest in a row per week & 21.65 & 24.85 & 3.20  & 0 \\
			Average number of train trips for each engineer per week & 2.72  & 2.57  & -0.15 & 0 \\
			Average number of driving hours for each engineer per week & 25.98 & 24.54 & -1.44 & 0 \\
			Average number of working hours for each engineer per week & 34.13 & 32.24 & -1.89 & 0 \\
			Percentage of trips driven from engineer's home base (\%) & 98.93 & 99.05 & 0.12  & 0 \\
			Average number of delayed trips per week & 13.40 & 14.74 & 1.34  & 0 \\
			Average number of delayed hours for each delayed trip & 0.80  & 0.79  & 0.00  & 0.0118 \\
			Average number of delayed hours for all trips & 0.17  & 0.19  & 0.02  & 0 \\
			\bottomrule
	\end{tabular}}
	\label{tbl_3city_buff_method2}%
\end{table}%

\section{Tablularized results for 2-city and 3-city instances obtained from Method 2: Direct Algorithm with $\beta=1.2$}\label{appendix_results_2/3city_method2}

\noindent The tables in this appendix have similar logic to tables in Appendix \ref{appendix_results_2/3city} and focus results obtained from the Direct Algorithm with $\beta=1.2$. 	
Table \ref{tbl_2city_summary_method2} summarizes the 2-city cost metrics obtained from the Direct Algorithm for all scenarios with $\beta=1.2$. Tables \ref{tbl_2city_length_method2}-\ref{tbl_2city_YN_method2} display the means and statistical test p-values  for the 2-city cost metrics as a function of call window lengths, maximum delayed hours, and whether to require engineers to go home by van, respectively.

Table \ref{tbl_3city_summary_method2} summarizes the 3-city cost metrics obtained from the Direct Algorithm for all scenarios with $\beta=1.2$. Tables \ref{tbl_3city_length_method2}-\ref{tbl_3city_YN_method2} display the means and statistical test p-values for the  3-city  cost metrics as a function of call window lengths, maximum delayed hours, and whether to require engineers to return home by van, respectively. 

\begin{landscape}
	
	\begin{table}[htbp]
		\centering
		\caption{Average results for 25 2-city instances obtained from Direct Algorithm for all scenarios with $\beta=1.2$}
		\resizebox{\columnwidth}{!}{\begin{tabular}{|c|c|c|c|c|c|c|c|c|c|c|c|c|c|c|}
				\toprule
				\multicolumn{1}{|p{4.145em}|}{Call window length (hr)} & \multicolumn{1}{p{4.88em}|}{Maximum delay (hr)} & \multicolumn{1}{p{4.145em}|}{Always take van home} & \multicolumn{1}{p{4.145em}|}{Avg. undercoverage rate (\%)} & \multicolumn{1}{p{4.145em}|}{Avg. no. working engineers} & \multicolumn{1}{p{4.145em}|}{Avg. no. engineers with at least one 48-hr rest in a row per week} & \multicolumn{1}{p{4.145em}|}{Avg. no. train trips for each engineer per week} & \multicolumn{1}{p{4.145em}|}{Avg. no. driving hours for each engineer per week} & \multicolumn{1}{p{4.145em}|}{Avg. no. working hours for each engineer per week} & \multicolumn{1}{p{4.145em}|}{Avg. no. trips driven by engineers from home base} & \multicolumn{1}{p{4.145em}|}{Avg. no. trips driven by engineers from away base} & \multicolumn{1}{p{4.145em}|}{Avg. no. delayed trips per week} & \multicolumn{1}{p{4.145em}|}{Avg. no. delayed hours for each delayed trip} & \multicolumn{1}{p{4.145em}|}{Avg. no. delayed hours for all trips} & \multicolumn{1}{p{4.97em}|}{Simulation runtime (sec)} \\
				\midrule
				\multirow{6}[6]{*}{6} & \multirow{2}[2]{*}{0} & Y     & 15.84 & 18.04 & 16.41 & 1.97  & 16.36 & 22.26 & 141.48 & 0.00  & 0.00  & 0.00  & 0.00  & 1219.72 \\
				&       & N     & 16.09 & 18.04 &  16.43 & 1.96  & 16.30 & 22.18 & 138.96 & 2.08  & 0.00  & 0.00  & 0.00  & 1450.64 \\
				\cmidrule{2-15}          & \multirow{2}[2]{*}{1} & Y     & 10.94 & 18.04 &  16.56 & 2.08  & 17.30 & 23.54 & 149.68 & 0.00  & 11.00 & 1.00  & 0.26  & 1404.18 \\
				&       & N     & 10.71 & 18.04 &  16.46 & 2.09  & 17.36 & 23.62 & 147.64 & 2.40  & 11.13 & 1.00  & 0.26  & 1395.60 \\
				\cmidrule{2-15}          & \multirow{2}[2]{*}{2} & Y     & 10.46 & 18.04 &  16.35 & 2.09  & 17.43 & 23.71 & 150.56 & 0.00  & 16.81 & 1.33  & 0.53  & 1543.78 \\
				&       & N     & 11.01 & 18.04 &  16.28 & 2.08  & 17.31 & 23.55 & 147.48 & 2.08  & 16.39 & 1.33  & 0.52  & 1620.52 \\
				\midrule
				\multirow{6}[6]{*}{10} & \multirow{2}[2]{*}{0} & Y     & 5.86  & 18.04 &  16.22 & 2.20  & 18.26 & 24.87 & 158.20 & 0.00  & 0.00  & 0.00  & 0.00  & 939.60 \\
				&       & N     & 6.05  & 18.04 &  16.19 & 2.20  & 18.26 & 24.85 & 153.72 & 4.16  & 0.00  & 0.00  & 0.00  & 854.38 \\
				\cmidrule{2-15}          & \multirow{2}[2]{*}{1} & Y     & 3.91  & 18.00 &  15.85 & 2.25  & 18.69 & 25.45 & 161.48 & 0.00  & 8.29  & 1.00  & 0.20  & 1121.20 \\
				&       & N     & 3.53  & 17.92 &  15.84 & 2.27  & 18.85 & 25.67 & 158.32 & 3.76  & 8.27  & 1.00  & 0.20  & 1154.96 \\
				\cmidrule{2-15}          & \multirow{2}[2]{*}{2} & Y     & 3.32  & 17.92 &  15.55 & 2.28  & 18.91 & 25.74 & 162.44 & 0.00  & 11.94 & 1.33  & 0.38  & 1104.69 \\
				&       & N     & 3.07  & 17.92 &  15.54 & 2.28  & 18.93 & 25.77 & 159.28 & 3.60  & 11.64 & 1.32  & 0.36  & 1195.25 \\
				\midrule
				\multirow{6}[6]{*}{12} & \multirow{2}[2]{*}{0} & Y     & 3.43  & 18.00 &  16.09 & 2.26  & 18.80 & 25.59 & 162.24 & 0.00  & 0.00  & 0.00  & 0.00  & 590.70 \\
				&       & N     & 3.39  & 17.92 &  15.82 & 2.28  & 18.88 & 25.71 & 157.92 & 4.40  & 0.00  & 0.00  & 0.00  & 714.90 \\
				\cmidrule{2-15}          & \multirow{2}[2]{*}{1} & Y     & 2.13  & 17.60 &  15.51 & 2.34  & 19.46 & 26.50 & 164.44 & 0.00  & 7.85  & 1.00  & 0.19  & 872.07 \\
				&       & N     & 2.25  & 17.80 &  15.46 & 2.32  & 19.25 & 26.19 & 160.20 & 4.04  & 7.69  & 1.00  & 0.18  & 1007.02 \\
				\cmidrule{2-15}          & \multirow{2}[2]{*}{2} & Y     & 2.05  & 17.84 &  15.36 & 2.32  & 19.23 & 26.19 & 164.60 & 0.00  & 11.31 & 1.34  & 0.36  & 1140.53 \\
				&       & N     & 1.70  & 17.68 &  15.29 & 2.35  & 19.47 & 26.50 & 160.88 & 4.32  & 11.02 & 1.34  & 0.35  & 1084.68 \\
				\midrule
				\multirow{6}[6]{*}{24} & \multirow{2}[2]{*}{0} & Y     & 5.39  & 17.88 &  16.19 & 2.23  & 18.56 & 25.26 & 158.96 & 0.00  & 0.00  & 0.00  & 0.00  & 2687.39 \\
				&       & N     & 4.90  & 17.88 &  16.21 & 2.24  & 18.66 & 25.39 & 157.64 & 2.16  & 0.00  & 0.00  & 0.00  & 2575.33 \\
				\cmidrule{2-15}          & \multirow{2}[2]{*}{1} & Y     & 4.88  & 17.72 &  16.11 & 2.26  & 18.81 & 25.60 & 159.80 & 0.00  & 6.42  & 1.00  & 0.15  & 2609.26 \\
				&       & N     & 4.73  & 17.76 &  15.92 & 2.26  & 18.78 & 25.56 & 157.64 & 2.44  & 6.39  & 1.00  & 0.15  & 2593.80 \\
				\cmidrule{2-15}          & \multirow{2}[2]{*}{2} & Y     & 4.68  & 17.76 &  16.03 & 2.26  & 18.80 & 25.59 & 160.12 & 0.00  & 7.33  & 1.26  & 0.22  & 4991.82 \\
				&       & N     & 4.67  & 17.76 &  15.98 & 2.26  & 18.79 & 25.58 & 157.44 & 2.72  & 7.54  & 1.25  & 0.22  & 6614.98 \\
				\bottomrule
		\end{tabular}}
		\label{tbl_2city_summary_method2}%
	\end{table}%

	\begin{table}[htbp]
		\centering
		\caption{Means and ANOVA p-values of costs with each call window length for 2-city instances obtained from Direct Algorithm for all scenarios with $\beta=1.2$}
		\resizebox{\columnwidth}{!}{\begin{tabular}{|l|c|c|c|c|c|c|}
				\hline
				\multicolumn{1}{|p{4.125em}|}{Cost}  & \multicolumn{1}{p{4.125em}|}{6-hr} & \multicolumn{1}{p{4.125em}|}{10-hr} & \multicolumn{1}{p{4.125em}|}{12-hr} & \multicolumn{1}{p{4.125em}|}{24-hr} & \multicolumn{1}{p{4.125em}|}{p-value} & \multicolumn{1}{p{4.125em}|}{Significant impact} \\ \hline
				Undercoverage rate (\%) & 12.51 & 4.29  & 2.49  & 4.88  & 0 & \checkmark \\
				Average number of working engineers & 18.04 & 17.97 & 17.81 & 17.79 & 0.4938 & \\
				Average number of engineers with at least one 48-hr rest in a row per week & 16.42 & 15.87 & 15.59 & 16.07 & 0 & \checkmark \\
				Average number of train trips for each engineer per week & 2.04  & 2.25  & 2.31  & 2.25  & 0 & \checkmark \\
				Average number of driving hours for each engineer per week & 17.01 & 18.65 & 19.18 & 18.73 & 0 & \checkmark \\
				Average number of working hours for each engineer per week & 23.14 & 25.39 & 26.12 & 25.50 & 0 & \checkmark \\
				Percentage of trips driven from engineer's home base (\%) & 99.26 & 98.80 & 98.70 & 99.24 & 0 & \checkmark \\
				Average number of delayed trips per week & 9.22  & 6.69  & 6.31  & 4.61  & 0 & \checkmark \\
				Average number of delayed hours for each delayed trip & 0.78  & 0.77  & 0.78  & 0.75  & 0.9697 &  \\
				Average number of delayed hours for all trips & 0.26  & 0.19  & 0.18  & 0.12  & 0 & \checkmark \\
				\bottomrule
		\end{tabular}}
		\label{tbl_2city_length_method2}%
	\end{table}%

	\begin{table}[htbp]
		\centering
		\caption{Means and ANOVA p-values of costs with each upper limit on the number of delayed hours for 2-city instances obtained from Direct Algorithm for all scenarios with $\beta=1.2$}
		\resizebox{\columnwidth}{!}{\begin{tabular}{|l|c|c|c|c|c|}
				\hline
				Cost  & \multicolumn{1}{p{4.125em}|}{0-hr} & \multicolumn{1}{p{4.125em}|}{1-hr} & \multicolumn{1}{p{4.125em}|}{2-hr} & \multicolumn{1}{p{4.125em}|}{p-value} & \multicolumn{1}{p{4.125em}|}{Significant impact} \\ \hline
				Undercoverage rate (\%) & 7.62  & 5.39  & 5.12  & 0 & \checkmark \\
				Average number of working engineers & 17.98 & 17.86 & 17.87 & 0.7299 & \\
				Average number of engineers with at least one 48-hr rest in a row per week & 16.20 & 15.96 & 15.80 & 0 & \checkmark \\
				Average number of train trips for each engineer per week & 2.17  & 2.24  & 2.24  & 0 & \checkmark \\
				Average number of driving hours for each engineer per week & 18.01 & 18.56 & 18.61 & 0 & \checkmark \\
				Average number of working hours for each engineer per week & 24.51 & 25.27 & 25.33 & 0 & \checkmark \\
				Percentage of trips driven from engineer's home base (\%) & 98.98 & 99.01 & 99.01 & 0.9630 & \\
				Average number of delayed trips per week & 0.00  & 8.38  & 11.75 & 0 & \checkmark \\
				Average number of delayed hours for each delayed trip & 0.00  & 1.00  & 1.31  & 0 & \checkmark \\
				Average number of delayed hours for all trips & 0.00  & 0.20  & 0.37  & 0 & \checkmark \\
				\bottomrule
		\end{tabular}}
		\label{tbl_2city_delayhour_method2}%
	\end{table}%

	\begin{table}[htbp]
		\centering
		\caption{Means and paired t-test p-values of costs under conditions whether to always require engineers to return home by van (Y), or to allow engineers to drive a train home (N) for 2-city instances obtained from Direct Algorithm for all scenarios with $\beta=1.2$}
		\resizebox{\columnwidth}{!}{\begin{tabular}{|l|c|c|c|c|}
				\hline
				Cost  & \multicolumn{1}{p{4.125em}|}{Y (van home)} & \multicolumn{1}{p{4.125em}|}{N (allow driving home)} & \multicolumn{1}{p{4.125em}|}{p-value} & \multicolumn{1}{p{4.125em}|}{Significant impact} \\ \hline
				Undercoverage rate (\%) & 6.07  & 6.01  & 0.4586 &  \\
				Average number of working engineers & 17.91 & 17.90 & 0.7395 & \\
				Average number of engineers with at least one 48-hr rest in a row per week & 16.02 & 15.95 & 0.0112 & \checkmark \\
				Average number of train trips for each engineer per week & 2.21  & 2.22  & 0.5693 &  \\
				Average number of driving hours for each engineer per week & 18.38 & 18.40 & 0.4598 &  \\
				Average number of working hours for each engineer per week & 25.02 & 25.05 & 0.4823 &  \\
				Percentage of trips driven from engineer's home base (\%) & 100.00 & 98.00 & 0 & \checkmark \\
				Average number of delayed trips per week & 6.75  & 6.67  & 0.1725 &  \\
				Average number of delayed hours for each delayed trip & 0.77  & 0.77  & 0.3208 &  \\
				Average number of delayed hours for all trips & 0.1909  & 0.1876  & 0.0403 & \checkmark \\
				\bottomrule
		\end{tabular}}
		\label{tbl_2city_YN_method2}%
	\end{table}%

	\begin{table}[htbp]
		\centering
		\caption{Average results for  25 3-city instances obtained from Direct Algorithm for all scenarios with $\beta=1.2$}
		\resizebox{\columnwidth}{!}{\begin{tabular}{|c|c|c|c|c|c|c|c|c|c|c|c|c|c|c|}
				\toprule
				\multicolumn{1}{|p{4.145em}|}{Call window length (hr)} & \multicolumn{1}{p{4.97em}|}{Maximum delay (hr)} & \multicolumn{1}{p{4.145em}|}{Always take van home} & \multicolumn{1}{p{4.145em}|}{Avg. undercoverage rate (\%)} & \multicolumn{1}{p{4.145em}|}{Avg. no. working engineers} & \multicolumn{1}{p{4.145em}|}{Avg. no. engineers with at least one 48-hr rest in a row per week} & \multicolumn{1}{p{4.145em}|}{Avg. no. train trips for each engineer per week} & \multicolumn{1}{p{4.145em}|}{Avg. no. driving hours for each engineer per week} & \multicolumn{1}{p{4.145em}|}{Avg. no. working hours for each engineer per week} & \multicolumn{1}{p{4.145em}|}{Avg. no. trips driven by engineers from home base} & \multicolumn{1}{p{4.145em}|}{Avg. no. trips driven by engineers from away base} & \multicolumn{1}{p{4.145em}|}{Avg. no. delayed trips per week} & \multicolumn{1}{p{4.145em}|}{Avg. no. delayed hours for each delayed trip} & \multicolumn{1}{p{4.145em}|}{Avg. no. delayed hours for all trips} & \multicolumn{1}{p{5.62em}|}{Simulation runtime (sec)} \\
				\midrule
				\multirow{6}[6]{*}{6} & \multirow{2}[2]{*}{0} & Y     & 37.18 & 26.72 & 25.04 & 2.24  & 21.38 & 28.09 & 238.76 & 0.00  & 0.00  & 0.00  & 0.00  & 663.19 \\
				&       & N     & 36.39 & 26.72 &  24.99 & 2.27  & 21.62 & 28.41 & 236.56 & 5.20  & 0.00  & 0.00  & 0.00  & 581.66 \\
				\cmidrule{2-15}          & \multirow{2}[2]{*}{1} & Y     & 32.83 & 26.72 &  25.16 & 2.39  & 22.85 & 30.03 & 255.24 & 0.00  & 22.69 & 1.00  & 0.24  & 853.12 \\
				&       & N     & 32.54 & 26.72 &  25.10 & 2.40  & 22.96 & 30.17 & 251.72 & 4.64  & 22.27 & 1.00  & 0.23  & 875.47 \\
				\cmidrule{2-15}          & \multirow{2}[2]{*}{2} & Y     & 32.85 & 26.72 &  25.20 & 2.39  & 22.82 & 30.00 & 255.20 & 0.00  & 31.80 & 1.39  & 0.46  & 8060.56 \\
				&       & N     & 32.05 & 26.72 &  25.09 & 2.42  & 23.09 & 30.34 & 253.20 & 4.96  & 31.52 & 1.39  & 0.46  & 1656.66 \\
				\midrule
				\multirow{6}[6]{*}{10} & \multirow{2}[2]{*}{0} & Y     & 28.88 & 26.68 &  24.77 & 2.54  & 24.23 & 31.84 & 270.32 & 0.00  & 0.00  & 0.00  & 0.00  & 511.74 \\
				&       & N     & 27.56 & 26.72 &  24.86 & 2.58  & 24.67 & 32.41 & 268.12 & 7.12  & 0.00  & 0.00  & 0.00  & 505.26 \\
				\cmidrule{2-15}          & \multirow{2}[2]{*}{1} & Y     & 25.46 & 26.64 &  24.89 & 2.66  & 25.42 & 33.40 & 283.20 & 0.00  & 18.71 & 1.00  & 0.20  & 643.95 \\
				&       & N     & 24.98 & 26.60 &  24.84 & 2.68  & 25.62 & 33.67 & 278.68 & 6.36  & 19.03 & 1.00  & 0.20  & 650.34 \\
				\cmidrule{2-15}          & \multirow{2}[2]{*}{2} & Y     & 24.69 & 26.64 &  24.70 & 2.69  & 25.73 & 33.80 & 286.12 & 0.00  & 27.50 & 1.41  & 0.41  & 854.05 \\
				&       & N     & 23.86 & 26.64 &  24.63 & 2.72  & 26.00 & 34.16 & 283.40 & 5.96  & 27.67 & 1.41  & 0.41  & 838.19 \\
				\midrule
				\multirow{6}[6]{*}{12} & \multirow{2}[2]{*}{0} & Y     & 25.96 & 26.64 &  24.79 & 2.64  & 25.30 & 33.23 & 281.32 & 0.00  & 0.00  & 0.00  & 0.00  & 688.27 \\
				&       & N     & 25.10 & 26.60 &  24.79 & 2.68  & 25.57 & 33.61 & 277.20 & 7.40  & 0.00  & 0.00  & 0.00  & 499.10 \\
				\cmidrule{2-15}          & \multirow{2}[2]{*}{1} & Y     & 23.78 & 26.64 &  24.56 & 2.72  & 25.98 & 34.14 & 289.64 & 0.00  & 18.27 & 1.00  & 0.19  & 641.62 \\
				&       & N     & 23.05 & 26.60 &  24.62 & 2.75  & 26.29 & 34.55 & 285.08 & 7.24  & 17.39 & 1.00  & 0.18  & 770.59 \\
				\cmidrule{2-15}          & \multirow{2}[2]{*}{2} & Y     & 22.82 & 26.56 &  24.44 & 2.77  & 26.37 & 34.67 & 293.24 & 0.00  & 26.07 & 1.42  & 0.39  & 906.40 \\
				&       & N     & 22.18 & 26.60 &  24.33 & 2.78  & 26.61 & 34.96 & 288.76 & 6.96  & 25.66 & 1.43  & 0.38  & 795.94 \\
				\midrule
				\multirow{6}[6]{*}{24} & \multirow{2}[2]{*}{0} & Y     & 29.39 & 26.56 &  25.10 & 2.53  & 24.16 & 31.75 & 268.20 & 0.00  & 0.00  & 0.00  & 0.00  & 11577.54 \\
				&       & N     & 29.30 & 26.48 &  25.07 & 2.54  & 24.28 & 31.90 & 266.56 & 1.96  & 0.00  & 0.00  & 0.00  & 2356.42 \\
				\cmidrule{2-15}          & \multirow{2}[2]{*}{1} & Y     & 29.09 & 26.52 &  24.99 & 2.54  & 24.34 & 31.97 & 269.40 & 0.00  & 14.14 & 1.00  & 0.15  & 7085.60 \\
				&       & N     & 28.72 & 26.48 &  24.82 & 2.56  & 24.51 & 32.20 & 268.24 & 2.60  & 14.08 & 1.00  & 0.15  & 2744.23 \\
				\cmidrule{2-15}          & \multirow{2}[2]{*}{2} & Y     & 29.01 & 26.36 &  24.83 & 2.56  & 24.49 & 32.17 & 269.64 & 0.00  & 18.39 & 1.31  & 0.25  & 5349.05 \\
				&       & N     & 28.36 & 26.48 &  24.85 & 2.57  & 24.61 & 32.33 & 269.36 & 2.76  & 18.46 & 1.32  & 0.25  & 4453.93 \\
				\bottomrule
		\end{tabular}}
		\label{tbl_3city_summary_method2}%
	\end{table}%

	\begin{table}[htbp]
		\centering
		\caption{Means and ANOVA p-values of costs with each call window length for 3-city instances obtained from Direct Algorithm for all scenarios with $\beta=1.2$}
		\resizebox{\columnwidth}{!}{\begin{tabular}{|l|c|c|c|c|c|c|}
				\toprule
				\multicolumn{1}{|p{4.125em}|}{Cost}  & \multicolumn{1}{p{4.125em}|}{6-hr} & \multicolumn{1}{p{4.125em}|}{10-hr} & \multicolumn{1}{p{4.125em}|}{12-hr} & \multicolumn{1}{p{4.125em}|}{24-hr} & \multicolumn{1}{p{4.125em}|}{p-value} & \multicolumn{1}{p{4.125em}|}{Significant impact} \\
				\midrule
				Undercoverage rate (\%) & 33.97 & 25.91 & 23.81 & 28.98 & 0 & \checkmark\\
				Average number of working engineers & 26.72 & 26.65 & 26.61 & 26.48 & 0.7203 & \\
				Average number of engineers with at least one 48-hr rest in a row per week & 25.10 & 24.78 & 24.59 & 24.94 & 0.0221 & \checkmark\\
				Average number of train trips for each engineer per week & 2.35  & 2.65  & 2.72  & 2.55  & 0 & \checkmark\\
				Average number of driving hours for each engineer per week & 22.45 & 25.28 & 26.02 & 24.40 & 0 & \checkmark\\
				Average number of working hours for each engineer per week & 29.51 & 33.22 & 34.19 & 32.05 & 0 & \checkmark\\
				Percentage of trips driven from engineer's home base (\%) & 99.03 & 98.87 & 98.77 & 99.55 & 0 & \checkmark\\
				Average number of delayed trips per week & 18.05 & 15.49 & 14.57 & 10.85 & 0 & \checkmark\\
				Average number of delayed hours for each delayed trip & 0.80  & 0.80  & 0.81  & 0.77  & 0.9460 &  \\
				Average number of delayed hours for all trips & 0.23  & 0.20  & 0.19  & 0.13  & 0 & \checkmark\\
				\bottomrule
		\end{tabular}}
		\label{tbl_3city_length_method2}%
	\end{table}%

	\begin{table}[htbp]
		\centering
		\caption{Means and ANOVA p-values of costs with each upper limit on the number of delayed hours for 3-city instances obtained from Direct Algorithm for all scenarios with $\beta=1.2$}
		\resizebox{\columnwidth}{!}{
			\begin{tabular}{|l|c|c|c|c|c|}
				\toprule
				Cost  & \multicolumn{1}{p{4.125em}|}{0-hr} & \multicolumn{1}{p{4.125em}|}{1-hr} & \multicolumn{1}{p{4.125em}|}{2-hr} & \multicolumn{1}{p{4.125em}|}{p-value} & \multicolumn{1}{p{4.125em}|}{Significant impact}  \\
				\midrule
				Undercoverage rate (\%) & 29.97 & 27.56 & 26.98 & 0 & \checkmark \\
				Average number of working engineers & 26.64 & 26.62 & 26.59 & 0.9645 & \\
				Average number of engineers with at least one 48-hr rest in a row per week & 24.93 & 24.87 & 24.76 & 0.5203 &  \\
				Average number of train trips for each engineer per week & 2.50  & 2.59  & 2.61  & 0 & \checkmark \\
				Average number of driving hours for each engineer per week & 23.90 & 24.75 & 24.96 & 0 & \checkmark \\
				Average number of working hours for each engineer per week & 31.41 & 32.52 & 32.80 & 0 & \checkmark \\
				Percentage of trips driven from engineer's home base (\%) & 99.00 & 99.07 & 99.09 & 0.7160 &  \\
				Average number of delayed trips per week & 0.00  & 18.32 & 25.88 & 0 & \checkmark \\
				Average number of delayed hours for each delayed trip & 0.00  & 1.00  & 1.38  & 0    & \checkmark \\
				Average number of delayed hours for all trips & 0.00  & 0.19  & 0.38  & 0 & \checkmark \\
				\bottomrule
		\end{tabular}}
		\label{tbl_3city_delayhour_method2}%
	\end{table}%

	\begin{table}[htbp]
		\centering
		\caption{Means and paired t-test p-values of costs under conditions whether to always require engineers to return home by van (Y), or to allow engineers to drive a train home (N) for 3-city instances obtained from Direct Algorithm for all scenarios with $\beta=1.2$}
		\resizebox{\columnwidth}{!}{\begin{tabular}{|l|c|c|c|c|}
				\hline
				Cost  & \multicolumn{1}{p{4.125em}|}{Y (van home)} & \multicolumn{1}{p{4.125em}|}{N (allow driving home)} & \multicolumn{1}{p{4.125em}|}{p-value} & \multicolumn{1}{p{4.125em}|}{Significant impact} \\
				\midrule
				Undercoverage rate (\%) & 28.49 & 27.84 & 0 & \checkmark \\
				Average number of working engineers & 26.62 & 26.61 & 0.7820 & \\
				Average number of engineers with at least one 48-hr rest in a row per week & 24.87 & 24.83 & 0.1354 &  \\
				Average number of train trips for each engineer per week & 2.56  & 2.58  & 0 & \checkmark \\
				Average number of driving hours for each engineer per week & 24.42 & 24.65 & 0 & \checkmark \\
				Average number of working hours for each engineer per week & 32.09 & 32.39 & 0 & \checkmark \\
				Percentage of trips driven from engineer's home base (\%) & 100.00 & 98.11 & 0 & \checkmark \\
				Average number of delayed trips per week & 14.80 & 14.67 & 0.1118 &  \\
				Average number of delayed hours for each delayed trip & 0.79  & 0.80  & 0.2402 &  \\
				Average number of delayed hours for all trips & 0.19  & 0.19  & 0.2972 &  \\
				\bottomrule
		\end{tabular}}
		\label{tbl_3city_YN_method2}%
	\end{table}%

\end{landscape}

\end{appendices}

\end{document}